\documentclass[twocolumn,english,showpacs,superscriptaddress,footinbib,bibnotes]{revtex4-1}  

\usepackage[version=3]{mhchem}

\usepackage{graphicx}
\usepackage[usenames,dvipsnames]{color}
\usepackage{dcolumn}
\usepackage{bbm}
\usepackage{bm}
\usepackage{amsmath}
\usepackage{amssymb}
\usepackage{times}
\usepackage{placeins}
\usepackage{hyperref}
\hypersetup{colorlinks=false}

\newcommand{\mc}[1]{\multicolumn{1}{c}{#1}}
\newcommand\Tstrut{\rule{0pt}{2.6ex}}
\newcommand\Bstrut{\rule[-0.9ex]{0pt}{0pt}}

\DeclareMathOperator\Tr{Tr}
\DeclareMathOperator\diag{diag}
\newcommand{\cdof}{\ensuremath \chi^{2}_{\text{dof}}}
\newcommand{\Id}[0]{\mathbbm{1}}
\newcommand{\dtau}{\ensuremath \Delta\tau}
\newcommand{\differentiald}[0]{\ensuremath \mathrm{d}}
\newcommand{\differentialD}[0]{\ensuremath D}

\begin{document}

\title{Competing Orders in a Nearly Antiferromagnetic Metal}

\author{Yoni Schattner}
\thanks{These authors have contributed equally to this work.}
\affiliation{Department of Condensed Matter Physics, The Weizmann Institute of Science, Rehovot, 76100, Israel}

\author{Max H. Gerlach}
\thanks{These authors have contributed equally to this work.}
\affiliation{Institute for Theoretical Physics, University of Cologne, 50937 Cologne, Germany}

\author{Simon Trebst}
\affiliation{Institute for Theoretical Physics, University of Cologne, 50937 Cologne, Germany}

\author{Erez Berg}
\affiliation{Department of Condensed Matter Physics, The Weizmann Institute of Science, Rehovot, 76100, Israel}

\date{\today}

\begin{abstract}
  We study the onset of spin-density wave order in itinerant electron
  systems via a two-dimensional lattice model amenable to numerically
  exact, sign-problem-free determinantal quantum Monte Carlo
  simulations.  The finite-temperature phase diagram of the model
  reveals a dome-shaped $d$-wave superconducting phase near the
  magnetic quantum phase transition.  Above the critical
  superconducting temperature, we observe an extended fluctuation
  regime, which manifests itself in the opening of a gap in the
  electronic density of states and an enhanced diamagnetic response.
  While charge density wave fluctuations are moderately enhanced in
  the proximity of the magnetic quantum phase transition, they remain
  short-ranged.  The striking similarity of our results to the
  phenomenology of many unconventional superconductors points a way to
  a microscopic understanding of such strongly coupled systems in a
  controlled manner.
\end{abstract}


\maketitle


A common feature of many strongly correlated metals, such as the
cuprates, the Fe-based superconductors, heavy-fermion compounds, and
organic superconductors, is the close proximity of unconventional
superconductivity (SC) and spin density wave (SDW) order in their
phase diagrams. This suggests that there is a common, universal
mechanism at work behind both phenomena~\cite{Scalapino2012note}. In
some of these systems, additional types of competing or coexisting
orders appear upon suppressing the SDW order, such as nematic,
charge-density wave (CDW), or possibly also pair density wave (PDW)
order.  Such a complex interplay between multiple types of electronic
order, with comparable onset temperature scales, is a recurring theme
in strongly correlated systems~\cite{Fradkin2015}.

These findings call for a detailed understanding of the physics of
{\em metals on the verge of an SDW transition}. It has long been
proposed that nearly--critical antiferromagnetic fluctuations can
mediate unconventional superconductivity~\cite{Scalapino1986,
  monthoux1991toward}. Many studies have focused on the universal
properties of an antiferromagnetic quantum critical point (QCP) in a
metal~\cite{Hertz1976, Millis1993, Abanov2000, Abanov2004, Abanov2003,
  Lohneysen2007,Metlitski2010a}. In particular, it has been proposed
that superconductivity is anomalously enhanced at the magnetic
QCP~\cite{Abanov2001Incoherent, abanov2008gap, Metlitski2010b,
  Wang2013a}. The same antiferromagnetic interaction can enhance other
subsidiary orders, such as CDW~\cite{Metlitski2010b, Efetov2013,
  Wang2014} or PDW~\cite{wang2015interplay, wang2015coexistence}. Near
the QCP, an approximate 
symmetry relating the SC and density wave order may
emerge~\cite{Metlitski2010b}. The resulting multi-component order
parameter would have a substantial fluctuation regime, proposed as the
origin of the ``pseudogap'' observed in the cuprates~\cite{Efetov2013,
  hayward2014angular,Meier2014,Pepin2014}. A deep minimum in the
penetration depth of the SC at low temperature, seen in the iron-based
SC $\mathrm{BaFe_2(As_{1-x}P_x)_2}$~\cite{Hashimoto2012}, has been
proposed as a generic manifestation of the underlying
antiferromagnetic QCP~\cite{levchenko2013enhancement,Shibauchi2014}.

\begin{figure}[b]
  \centering
  \includegraphics[width=\linewidth]{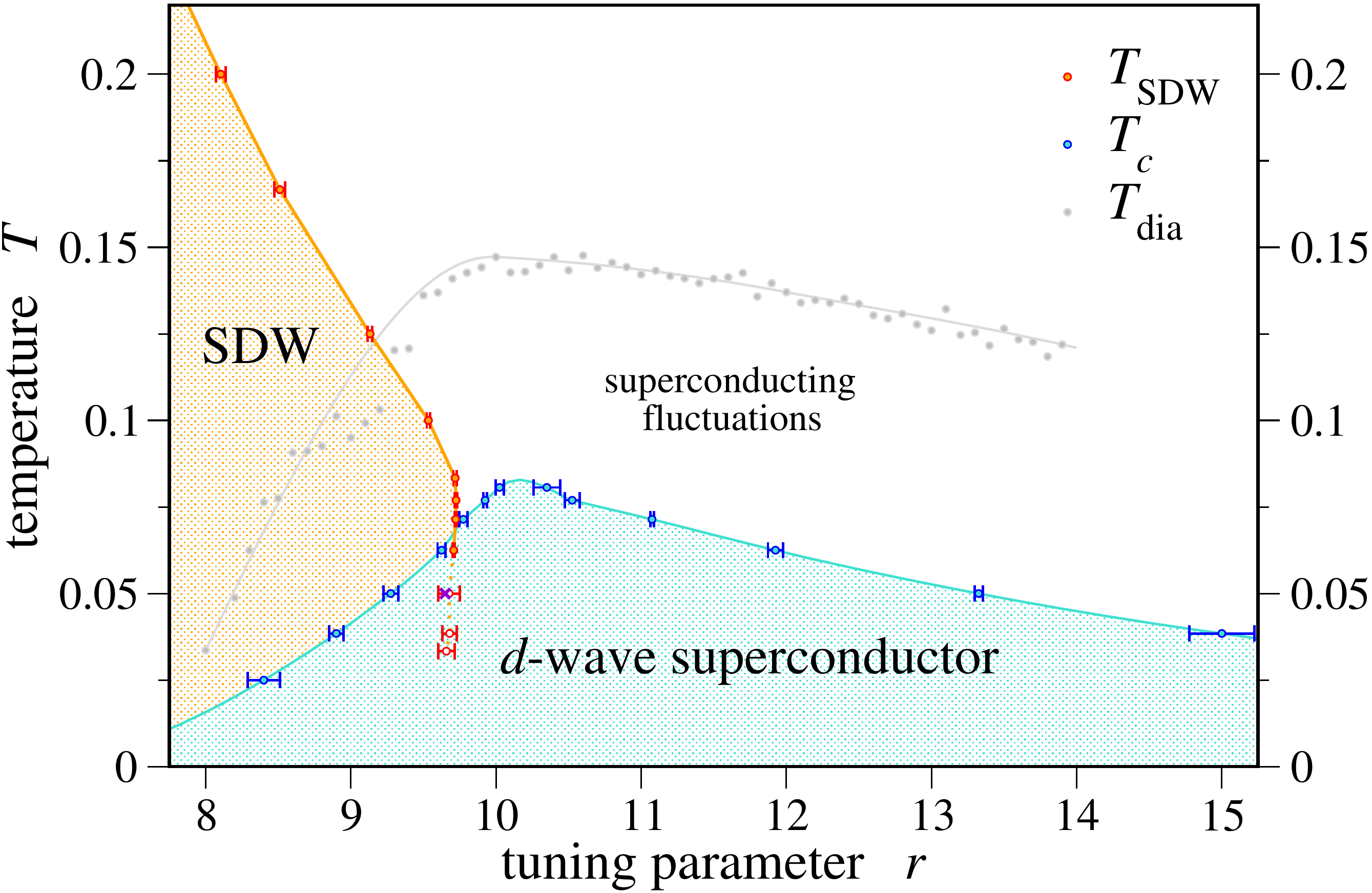}
  \caption{(Color online) Phase diagram of model~\eqref{eq:action}
    showing the transition temperature $T_{\mathrm{SDW}}$ to magnetic
    spin density wave (SDW) order, the superconducting $T_c$, and the
    onset of diamagnetism at $T_{\rm dia}$.  The solid lines indicate
    a Berezinskii-Kosterlitz-Thouless transition.  The SDW transition
    inside the SC dome, marked by a dashed line, possibly is a weakly
    first-order transition (see the main text).}
  \label{fig:phasediagram}
\end{figure}

Due to the strong coupling nature of the problem of a nearly
antiferromagnetic metal, obtaining analytically controlled solutions
has proven difficult.  In Ref.~\cite{Berg2012}, a two-dimensional
lattice model of a nearly-antiferromagnetic metal amenable to
sign-problem-free, determinantal quantum Monte-Carlo (DQMC)
simulations has been introduced. In this manuscript, we discuss the
finite-temperature phase diagram obtained by large scale simulations
of a closely related model.  Our simulations provide numerically
exact, unbiased results, which, when extrapolated to the thermodynamic
limit, are highly reminiscent of the behavior of many unconventional
superconductors. In the vicinity of the magnetic quantum phase
transition (see Fig.~\ref{fig:phasediagram}), we find a $d$-wave
superconducting dome with a maximum $T_c$ of the order of $E_{F}/30$,
where $E_{F}$ is the Fermi energy. Above $T_c$, there is a substantial
regime of strong superconducting fluctuations which is seen in a large
diamagnetic response and in a reduction of the tunneling density of
states. In the superconducting state we find a region of possible
coexistence with SDW order~\cite{comment-coexistence}.

In addition to SC order, we have examined CDW and PDW ordering
tendencies near the magnetic quantum phase transition (QPT). While the
CDW susceptibility shows a moderate enhancement in the vicinity of the
QPT, there is no sign of a near-degeneracy between the SC and CDW
order parameters as the QPT is approached. Finally, the
low-temperature superfluid density is found to vary smoothly through
the SC dome, similar to the behavior seen in the Co
doped~\cite{luan2011local} and unlike the P doped~\cite{Hashimoto2012}
$\mathrm{BaFe_2 As_2}$.

{\em Model.--} Our lattice model consists of two flavors of
spin--$\frac{1}{2}$ fermions, $\psi_x$ and $\psi_y$, coupled to an SDW
order parameter $\vec\varphi$. We set the magnetic ordering wavevector
to $\mathbf{Q} =(\pi,\pi)$. We assume that the SDW order parameter has
an {\em easy-plane} character, and restrict the order parameter
$\vec \varphi$ to lie in the $XY$ plane. Using an $O(2)$ rather than
$O(3)$ order parameter (as in Ref.~\cite{Berg2012}) gives rise to a
{\em finite-temperature} SDW phase transition of
Berezinskii-Kosterlitz-Thouless (BKT) character and, on a more
technical level, allows for higher numerical efficiency.

The action is $S=S_F + S_{\varphi} =
\int_0^{\beta}\differentiald\tau(L_F+L_{\varphi})$ with
\begin{align}
  \label{eq:action}
  L_F ={}& \sum_{\substack{i,j,s\\ \alpha = x,y}}
  \psi^{\dagger}_{\alpha i s} \left[ (\partial_{\tau}
  -\mu)\delta_{ij} - t_{\alpha ij} \right]
  \psi_{\alpha j s} \notag\\
  {}& + \lambda \sum_{\substack{
      i, s,s'}} e^{i \mathbf{Q \cdot r}_i}                
      [\vec{s} \cdot
      \vec{\varphi}_i ]_{ss'} \psi^{\dagger}_{xis}
      \psi_{yis'}^{\vphantom{\dagger}} + \mathrm{h.c.}, \notag\\
  L_{\varphi} ={}& \frac{1}{2} \sum\limits_{i}
                   \frac{1}{c^2}
                   \left( \frac{\differentiald\vec{\varphi}_i}{\differentiald\tau} \right)^2
                   + \frac{1}{2} \sum\limits_{\left\langle i,j
                   \right\rangle} \left(
                   \vec{\varphi}_i - 
                   \vec{\varphi}_j\right)^2
                   \notag\\
         &+ \sum\limits_{i} \left[ \frac{r}{2}\vec{\varphi}_i^2 +
           \frac{u}{4} (\vec{\varphi}_i^2)^2 \right] .  
\end{align}
Here $i,j$ label the sites of a square lattice, $\alpha=x,y$ are
flavor indices, $s,s'=\uparrow,\downarrow$ are spin indices, and
$\vec{s}$ are Pauli matrices. $\tau$ denotes imaginary time and
$\beta=1/T$ the inverse temperature. The hopping amplitudes for the
$\psi_x$-fermions along the horizontal and vertical lattice directions
are $t_{x,h}=1$ and $t_{x,v}=0.5$, respectively, while for the
$\psi_y$-fermions $t_{y,h}=0.5$ and $t_{y, v}=1$.  Note that for this
choice of parameters the dispersion of the $\psi_x$ and $\psi_y$
fermions is quasi one-dimensional with the two bands related by a
$\pi/2$ rotation.  $r$ is a tuning parameter used to tune the system
to the vicinity of an SDW instability. In an experimental context, $r$
can be thought of as doping or pressure.  We set the chemical
potential to $\mu=0.5$, the quartic coupling to $u=1$, the Yukawa
coupling to $\lambda = 3$, and the bare bosonic velocity to $c=2$.

{\em Numerical simulations.--} We study model \eqref{eq:action} by
extensive DQMC~\cite{Blankenbecler1981,White1989,LohJr.1992,
  Assaad2008} simulations, which due to the two-flavor structure of
the model do not suffer from the sign problem \cite{Berg2012}.  The
simulations were performed with a single flux quantum threaded through
the system, which dramatically improves the approach to the
thermodynamic limit for metallic systems~\cite{Assaad2002a}.
Specifically, we choose a magnetic flux whose direction for fermionic
spin-flavor pairs $(x\uparrow,y\downarrow)$ is opposite to the one for
$(x\downarrow,y\uparrow)$ pairs -- a setup which avoids the
reappearance of a sign problem~\cite{Schattner2015,Supplemental}.  For
details of this procedure and other technical aspects of the DQMC
simulations and data analysis we refer to the extensive Supplemental
Material. We report results up to linear extent $L=14$ and
temperatures down to $T=0.025$.

{\em Phase diagram.--} Our main finding is the phase diagram of model
\eqref{eq:action} as shown in Fig.~\ref{fig:phasediagram}.  The system
displays a quasi-long-range ordered SDW phase, whose transition
temperature, $T_{\mathrm{SDW}}$, decreases upon increasing $r$.  In
the vicinity of the magnetic QPT where $T_{\mathrm{SDW}}$ collapses to
zero, we find a region with quasi-long range $d$-wave superconducting
order. The superconducting $T_c$ traces an asymmetric dome-like shape
as a function of $r$ and reaches a maximum of $T_c^{\rm max} \approx 0.08$
at $r_{\rm opt} \approx 10.4$.

\begin{figure}[t]
   \centering
   \includegraphics[width=\linewidth]{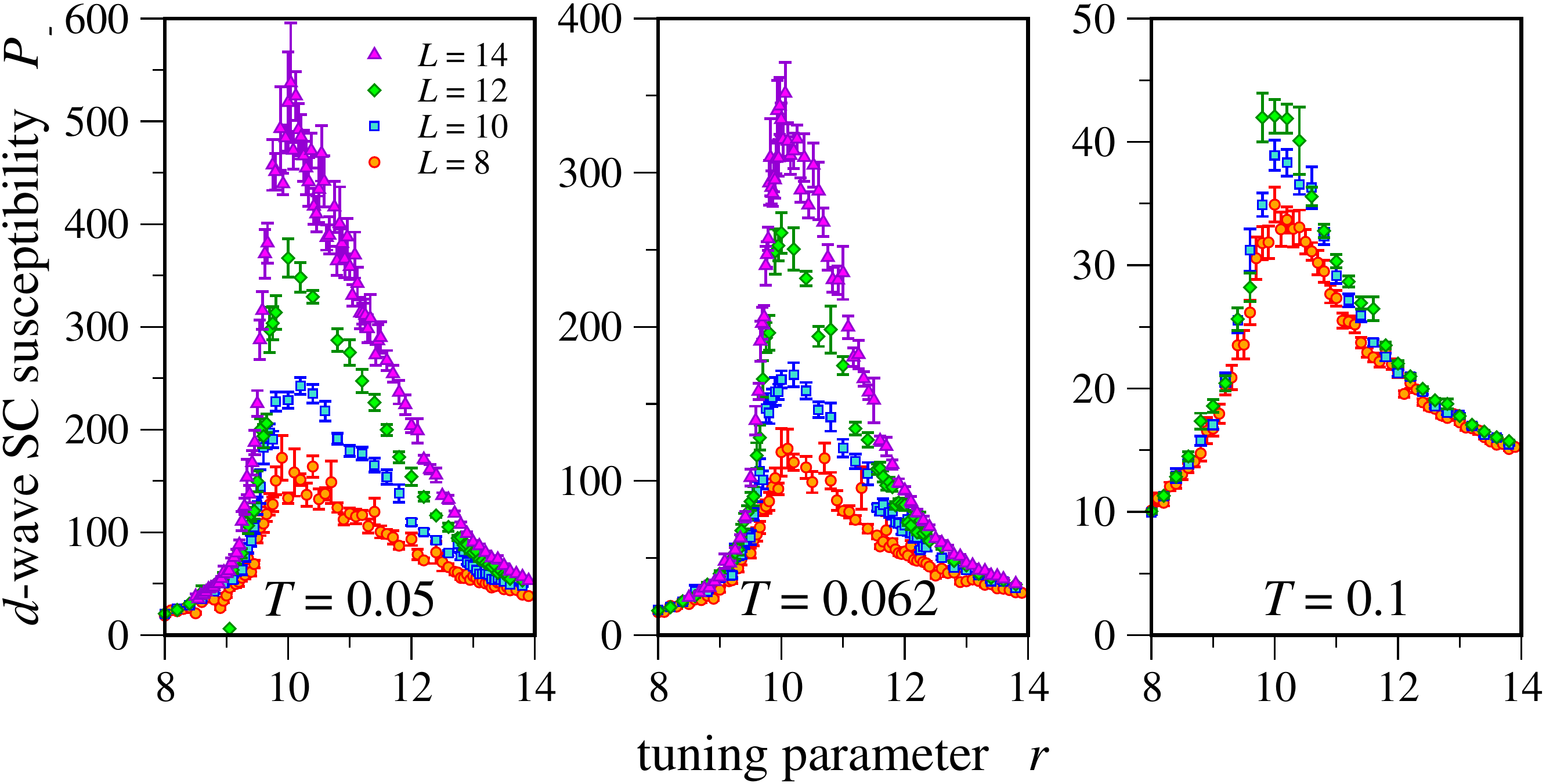}
   \caption{(Color online) $d$-wave superconducting susceptibility $P_{-}$ as a function of $r$ 
   		for different system sizes and temperatures.}
 \label{fig:Pminus}
\end{figure}

At sufficiently high temperatures, the antiferromagnetic transition is
consistent with BKT character. In this regime the SDW susceptibility
$\chi = \int d\tau \sum_i\langle \vec\varphi_i(\tau) \cdot
\vec\varphi_0(0)\rangle$ for different system sizes nicely follows the
expected scaling behavior $\chi \propto L^{2-\eta}$, with $\eta$
changing continuously as a function of $r$ and $T$, as illustrated in
the Supplemental Material. We identify $T_{\mathrm{SDW}}$ as the point
where we observe the BKT value $\eta=1/4$. At low temperatures,
$T \lesssim 0.05$ (i.e. within the SC region), the situation is less
clear with the numerical data starting to systematically deviate from
this scaling behavior.  In fact, there are indications that the
transition may become weakly first order at sufficiently low $T$, see
the discussion in the Supplemental Material. 

The SC transition is identified as
the point where the superfluid density obtains the universal, BKT value
$2T/\pi$ \cite{Paiva2004,Scalapino1993}, and is always consistent with BKT behavior.
The nature of the SC phase clearly reveals itself in the $d$-wave pairing susceptibility $P_{-} = \int d\tau \sum_i
\langle\Delta_{-}^{\dagger}(\mathbf{r}_i,\tau)\Delta_{-}(\mathbf{0},0)\rangle$ with 
$\Delta_{-}(\mathbf{r}_i)=\psi_{xi\uparrow}^{\dagger}\psi_{xi\downarrow}^\dagger - \psi_{yi\uparrow}^{\dagger}\psi_{yi\downarrow}^\dagger$, shown in Fig.~\ref{fig:Pminus}. At low temperatures $P_{-}$ is found to increase rapidly with system size, indicating that the SC phase has $d$-wave symmetry in the thermodynamic limit. The $s$-wave pairing susceptibility, in contrast, is found to be much smaller and system size independent \cite{Supplemental}.

\begin{figure}[t]
   \centering
   \includegraphics[width=\linewidth]{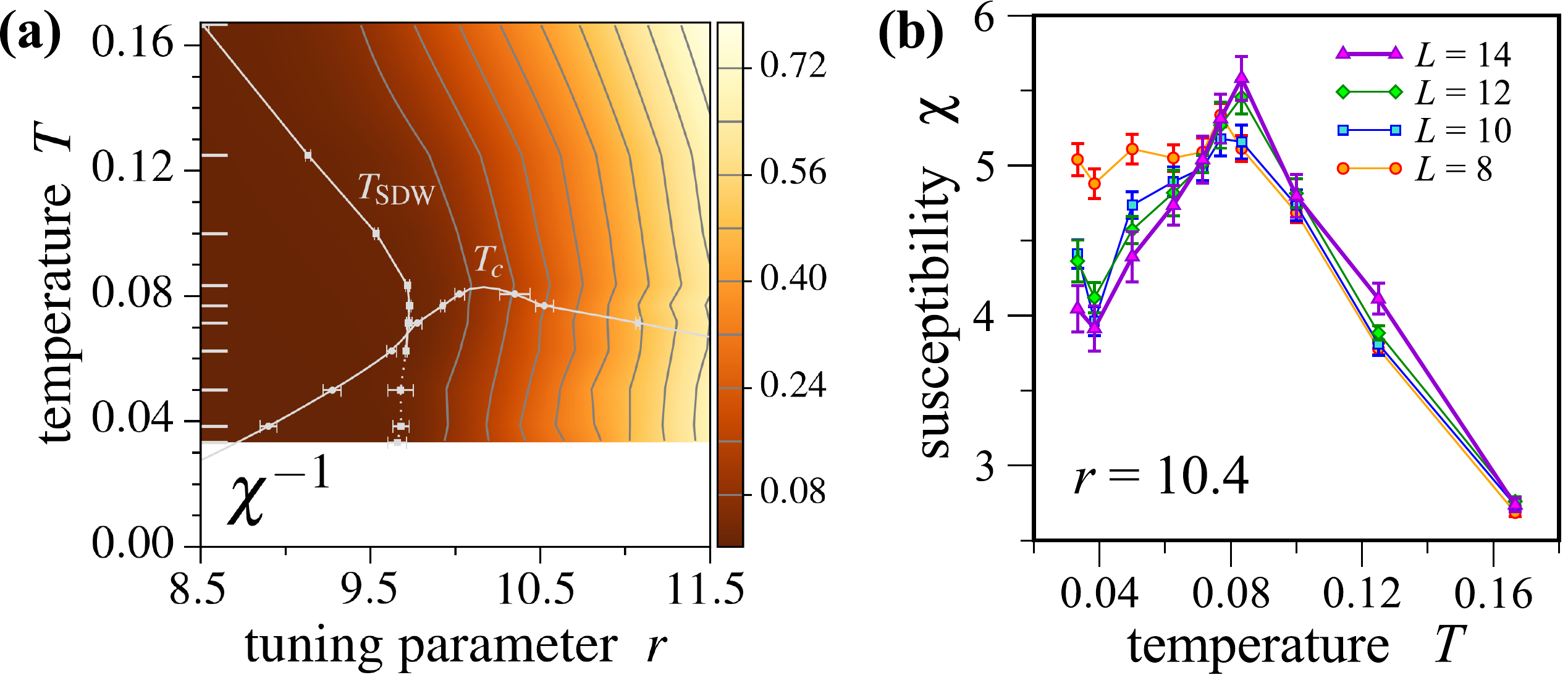}
   \caption{(Color online) (a) Inverse magnetic susceptibility across the phase diagram with
     the grey lines indicating contour lines. 
     We show data for
     $L=14$ at those temperatures
     indicated by ticks on the left inside of the plot and interpolate
     linearly between them.  (b) Magnetic susceptibility near the
     maximum of the superconducting dome ($r_{\rm opt}=10.4$) as a function of
     temperature for different system sizes.}
 \label{fig:chi_vs_T}
\end{figure}

A striking feature seen in the phase diagram is the ``bending'' of the magnetic phase boundary (indicated by $T_{\mathrm{SDW}}$) near the point where it crosses the superconducting dome. An even more pronounced back bending is  apparent in the magnetic susceptibility 
over a wide range of the tuning parameter $r$ as shown in Fig.~\ref{fig:chi_vs_T}(a).
Tracking the SDW susceptibility for fixed $r$, as shown in Fig.~\ref{fig:chi_vs_T}(b), one finds non-monotonic behavior with a maximum seen near $T_c$. 
Such a behavior has been predicted to arise from the competition between the two order parameters~\cite{moon2009competition}, and has been observed in certain unconventional superconductors, such as Ba$_{1-x}$Co$_x$Fe$_2$As$_2$~\cite{Ni2010}. 

\begin{figure}[t]
  \centering
  \includegraphics[width=\linewidth]{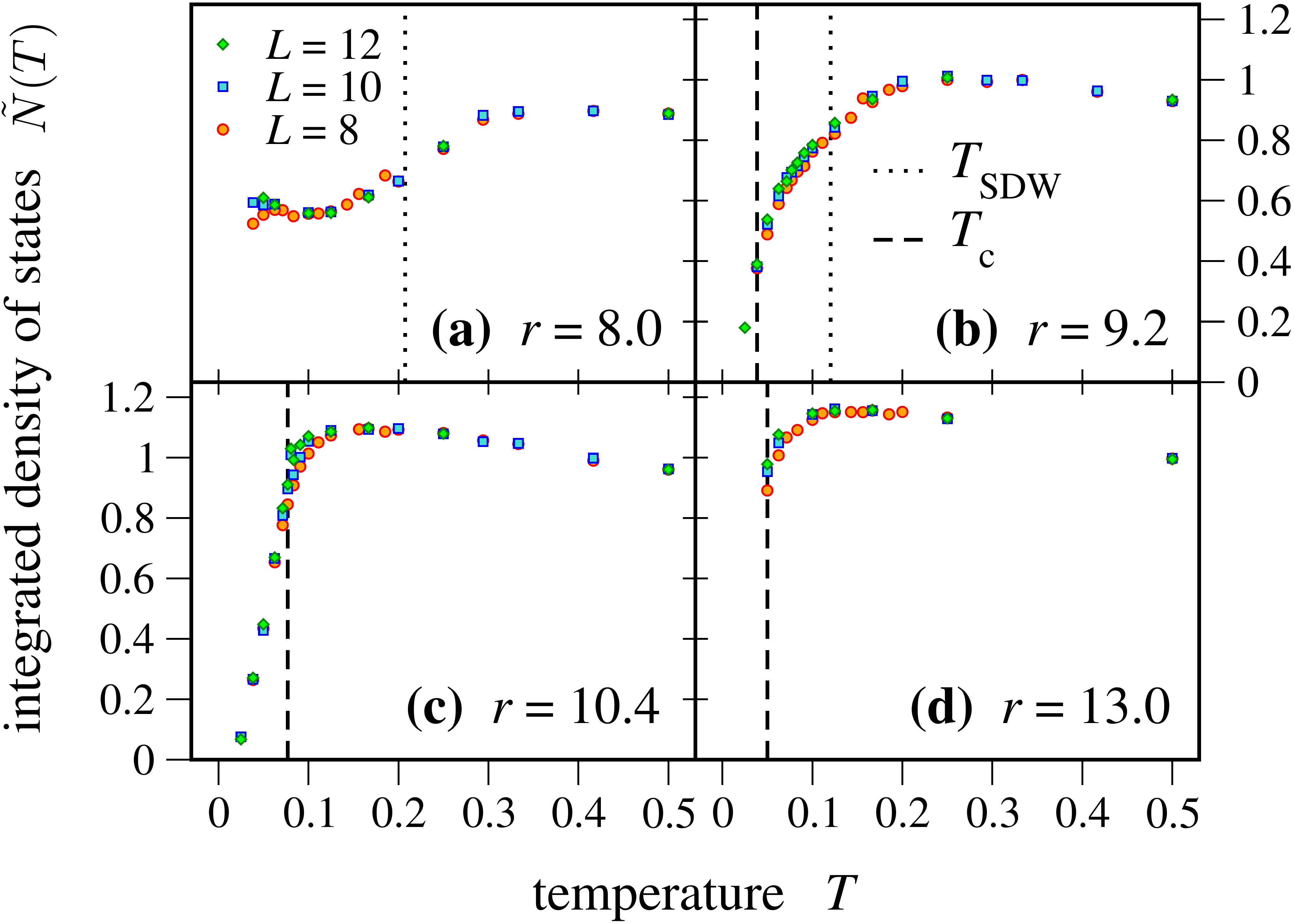}
  \caption{(Color online) The integrated density of states $\tilde N(T)$, as defined in \eqref{eq:dos}, versus temperature for various values of the tuning parameter $r$. The  dashed (dotted) lines indicate the location of the  SC (SDW) transition temperatures, respectively.}
  \label{fig:dos}
\end{figure}

In a finite range of temperatures above $T_c$, the orbital magnetic susceptibility is diamagnetic in sign (unlike the high temperature susceptibility, which is paramagnetic in our model), and its magnitude rapidly grows  with decreasing temperature. We identify this behavior as a signature of substantial finite-range superconducting fluctuations. The temperature where the orbital susceptibility changes sign, denoted by $T_{\rm dia}$, is indicated by the grey dots in Fig.~\ref{fig:phasediagram}. Over much of the phase diagram, $T_{\rm dia}$ roughly follows the shape of the superconducting dome, i.e. $T_{\rm dia} \propto T_c$. 
Another manifestation of finite-range superconducting fluctuations is the opening of a gap in the single-particle density of states $N(\omega, T)$ above $T_c$. While we cannot access $N(\omega,T)$ directly without performing an analytical continuation, we can use the relation~\cite{Trivedi1995}
\begin{equation} 
	\label{eq:dos} 
	\begin{split}
		\tilde{N}(T) &= \frac{1}{L^2 T} \textrm{Tr}\, G(\tau=\beta/2) \\ 
		&= \int_{-\infty}^\infty \frac{d\omega}{2 T\cosh(\beta \omega /2)} N(\omega,T),
	\end{split} 
\end{equation}
to extract information about the low-energy density of states, where $G$ is the imaginary time single-particle Green's function. Note that $\tilde{N}(T \rightarrow 0) =\pi N(\omega=0, T=0)$. This integrated density of states $\tilde{N}(T)$ is shown in Fig.~\ref{fig:dos}. In the SDW state, the behavior is consistent with a partial gapping of the Fermi surface (and corresponding suppression of the density of states), which commences slightly above the magnetic ordering temperature $T_{\mathrm{SDW}}$, see panels (a) and (b). A similar reduction of $\tilde N(T)$ is also found to set in above the superconducting $T_c$, see panels (b), (c) and (d).  Extrapolating $\tilde{N}(T)$ to $T=0$ indicates that the superconducting state is fully gapped.

{\em CDW and PDW susceptibilities.--}
To explore possible CDW and PDW instabilities, we turn to examine the susceptibilities of various density-wave orders near the magnetic quantum phase transition. Specifically, we define the CDW and PDW susceptibilities
\begin{equation}
\label{eq:susc}
\begin{split}
C_{\eta}(\mathbf{q}) &= \int d\tau\langle\tilde{\Delta}_{\eta}^{\dagger}(\mathbf{q},\tau)\tilde{\Delta}_{\eta}(\mathbf{q},0)\rangle,\\
P_{\eta}(\mathbf{q}) &= \int d\tau\langle\Delta_{\eta}^{\dagger}(\mathbf{q},\tau)\Delta_{\eta}(\mathbf{q},0)\rangle,
\end{split}
\end{equation}
where 
$\tilde{\Delta}_{\eta}(\mathbf{r}_i)=\sum_{s=\uparrow,\downarrow}\left(\psi_{xis}^{\dagger}\psi^{\phantom\dagger}_{xis} +\eta \psi_{yis}^{\dagger}\psi^{\phantom\dagger}_{yis}\right)$ 
and 
$\Delta_{\eta}(\mathbf{r}_i)=\psi_{xi\uparrow}^{\dagger}\psi_{xi\downarrow}^\dagger +\eta \psi_{yi\uparrow}^{\dagger}\psi_{yi\downarrow}^\dagger$ with
$\eta=\pm 1$. Note that under a $\pi/2$ rotation, associated with a rotation matrix $R_{\frac{\pi}{2}}$, we have $\Delta_{\eta}(\mathbf{r}) \rightarrow \eta \Delta_{\eta}(R_{\frac{\pi}{2}} \mathbf{r})$, i.e ${\Delta}_{-}$ 
has a $d$-wave (B$_{1g}$) character,
and similarly for $\tilde{\Delta}_-$.

\begin{figure}[b]
  \centering
  \includegraphics[width=\linewidth]{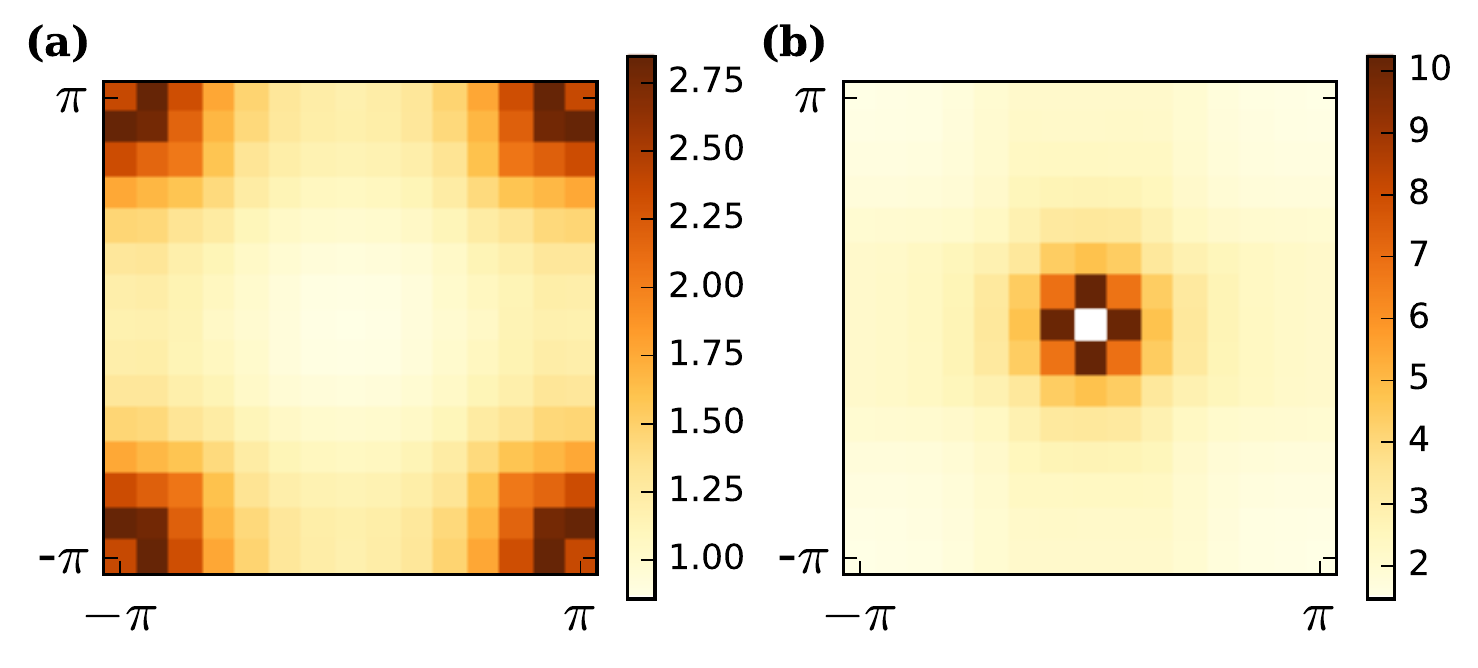}
  \caption{(Color online) (a) $d$-wave CDW and 
  		(b) $d$-wave PDW susceptibilities, as defined in Eq.~\eqref{eq:susc}, across the Brillouin zone. 
  		Shown here is data for $L=14$, $T=0.083$, and $r=10.4$. 
		The data point $P_-(q=0)$ (i.e. the uniform superconducting susceptiblity) has been excluded from the data (white square).}
  \label{fig:cdw_cmap}
\end{figure}

\begin{figure}[t]
  \centering
  \includegraphics[width=\linewidth]{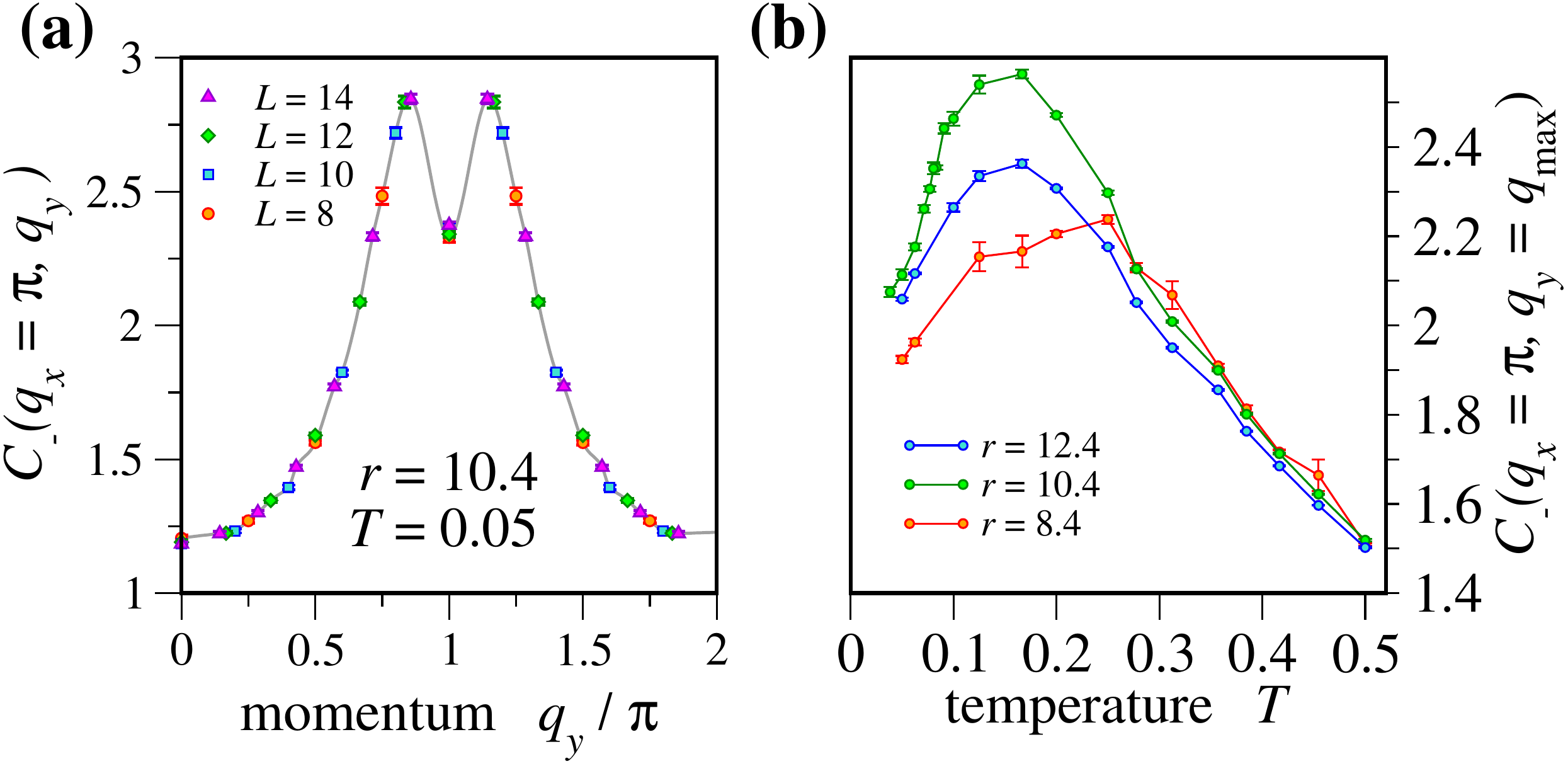}
  \caption{(Color online) (a) The $d$-wave CDW susceptibility versus momentum along the high-symmetry cut $\mathbf q=(\pi,q_y)$ for various system sizes. 
  					The solid line is a guide to the eye. 
				     (b) Temperature dependence of the CDW susceptibility at the optimal $q$ for multiple values of $r$.}
  \label{fig:cdw}
\end{figure}

In Fig.~\ref{fig:cdw_cmap} we show the momentum dependence of $C_-$ and $P_-$. $P_-$ is strongly peaked at $\mathbf{q}=0$ and does not display much structure at other momenta, indicating that there is no noticeable tendency towards PDW order. $P_+(\mathbf q=0)$ (not shown \cite{Supplemental}) 
is significantly smaller in amplitude, and also shows no structure at finite momenta.
$C_-$ is maximal in the vicinity of (but away from) $\mathbf q=(\pi,\pi)$. $C_+(\mathbf{q})$ (not shown \cite{Supplemental}) is qualitatively similar to $C_-(\mathbf{q})$, although its maximal value is approximately $3$ times lower.

Focusing on $C_-$, we show its momentum dependence along the high-symmetry cut $\mathbf{q} = (\pi, q_y)$  in Fig.~\ref{fig:cdw}(a).  The data taken from different system sizes collapses onto a single curve, suggesting that the 
CDW correlation length is sufficiently short such that results are representative of the thermodynamic limit. 
The temperature dependence of $C_-$ at the CDW wavevector $\mathbf{q}_{\mathrm{max}} = (\pi, q_{\mathrm{max}})$  where it is maximal ($q_{\mathrm{max}} \approx 0.92 \pi$) is shown in Fig.~\ref{fig:cdw}(b) for different values of $r$ on either side of the magnetic QPT. We find that $C_-(\mathbf{q}_{\mathrm{max}}, T)$ is maximal at a temperature close to $\mathrm{max}(T_c, T_{\mathrm{SDW}})$. This can be understood as a consequence of the reduction in the density of states due to the SC or SDW order.  
Across the entire phase diagram, the maximal CDW susceptibility is obtained at the value of $r$ close to the SDW QPT, where $T_c$ is also maximal. Note, however, that near $T_c$ the $d$-wave pairing susceptibility $P_{-}$ is at least an order of magnitude larger than the CDW susceptibility.

\begin{figure}[t]
\label{fig:rho_s}
  \centering
  \includegraphics[width=\linewidth]{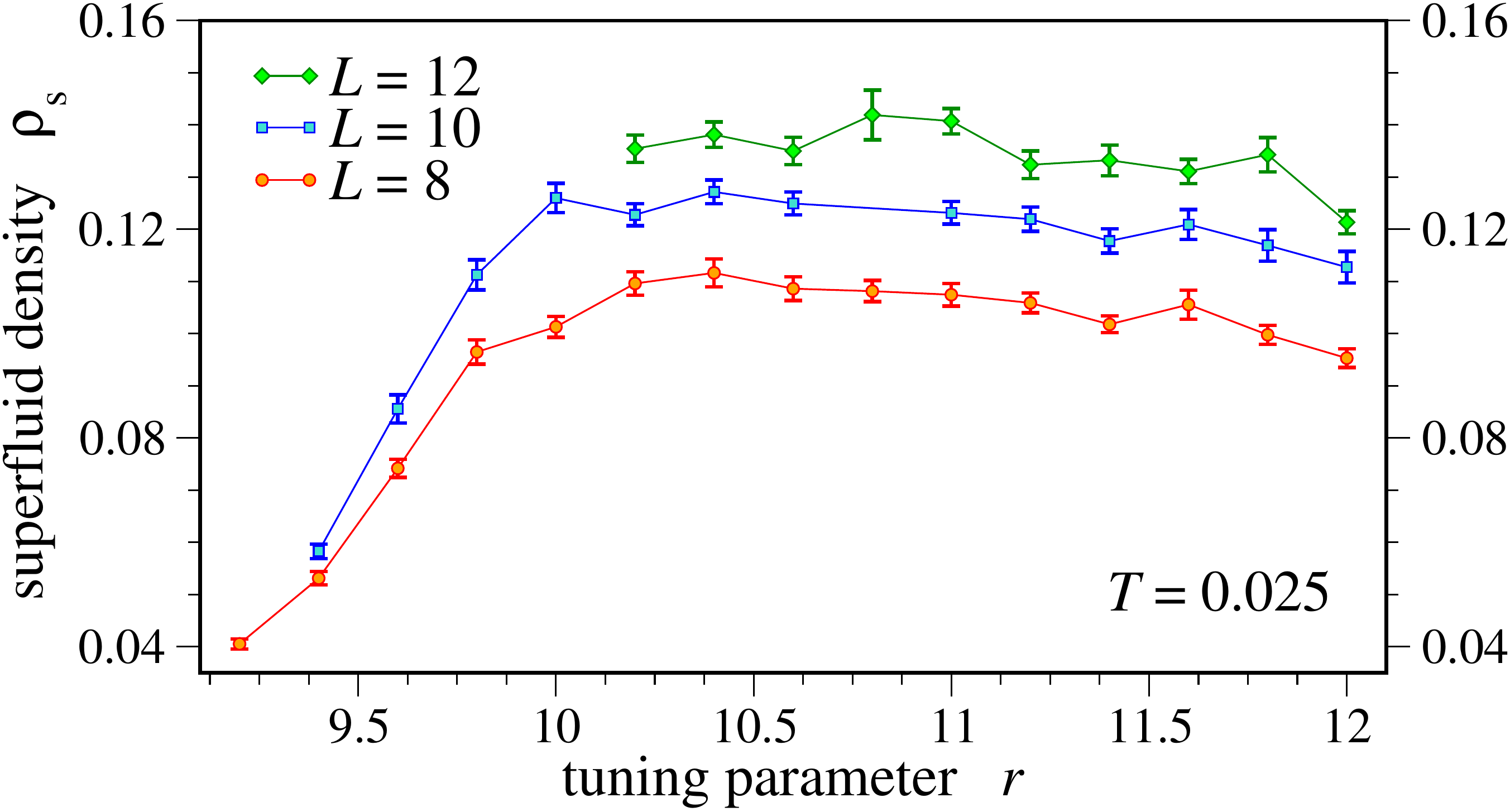}
  \caption{(Color online) The finite-size superfluid density $\rho_s(L)$ within the superconducting phase 
  		versus $r$ at $T=0.025$.}
  \label{fig:superfluiddensity}
\end{figure}

{\em Superfluid density.--}
Finally, we examine the low-temperature superfluid density across the phase diagram, proposed to exhibit a sharp minimum at a magnetic QCP~\cite{Shibauchi2014}. Figure~\ref{fig:superfluiddensity} shows the finite-size superfluid density  $\rho_s(L)$ \cite{Supplemental} along a cut through the superconducting dome at a fixed temperature, $T=0.025$. Notably, we find that inside the SC dome $\rho_s(L)$ is only weakly $r$-dependent, with no apparent minimum at $r_{\rm opt}\approx 10.4$.
This is consistent with the predictions of a field theoretical analysis~\cite{Chowdhury2013} and with the observed behavior in Ba$_{1-x}$Co$_x$Fe$_2$As$_2$~\cite{Ni2010}, and suggests that the sharp minimum observed in BaFe$_2$(As$_{1-x}$P$_x$)$_2$ may be of a different origin (see, e.g., Ref.~\cite{Chowdhury2015}).


{\em Discussion.--} 
The striking similarity between the phase diagram of our model (Fig.~\ref{fig:phasediagram}) and the phase diagrams of many unconventional superconductors, such as the iron-based SC, electron-doped cuprates or organic SC, strongly suggests that much of the essential physics in these systems is indeed captured within our model, as has been long anticipated~\cite{Scalapino2012note}. This encouraging result calls for further investigations of extensions of this basic model, designed to capture more material-specific features. For example, it would be interesting to consider a multiple component SDW order parameter (as in the pnictides), multiple bands, and additional competing order parameters. A  first step in this direction has been taken recently~\cite{Li2015a}.

Since models similar to the one studied here are frequently invoked to describe the phenomenology of the hole-doped cuprate superconductors, it is interesting to contrast the behavior seen in our model to that of the cuprates. Our model exhibits a gap in the single particle spectrum that precedes the phase transitions into the SC and SDW phases, as has been predicted for a nearly antiferromagnetic metal~\cite{Schmalian1999}. However, the onset temperature of the gap roughly follows the ordering temperature, and is never larger than about twice the corresponding transition temperature. In this sense, our results are different from the pseudogap phase of the cuprates. Our model also displays diamagnetic fluctuations with an onset temperature proportional to, and significantly above, $T_c$. Similar phenomena have been observed in the cuprates~\cite{Li2005,Li2010}.

In addition to unconventional superconductivity, our model exhibits an enhancement of CDW fluctuations with a $d$--wave form factor. However, the CDW susceptibility is only moderately enhanced compared to the expectation based on the non-interacting band structure. The quasi-one dimensional character of the dispersion of each fermion flavor enhances the CDW susceptibility, although at a non-zero chemical potential there is no ``perfect nesting''. See the Supplemental Material for a detailed comparison.  It seems that the interaction mediated by spin fluctuations is not sufficient, by itself, to promote strong CDW fluctuations. This is consistent with the conclusions of Refs.~\cite{sachdev2013bond, sau2014mean, Allais2014, Mishra2015} that additional, non-magnetic interactions are needed to stabilize a CDW phase.  

Finally, since in our model the magnetic phase transition occurs inside a superconducting phase, our results do not have a direct bearing on the question of metallic SDW quantum criticality. In addition, we found some indications that at low temperatures, the SDW transition may become weakly first order. Nevertheless, since $T_c$ is significantly smaller than $E_F$, one can still expect to see a substantial crossover regime above $T_c$ where the physics is dominated by an underlying ``avoided'' QCP. Indeed, we have preliminary indications that above $T_c$, the dynamic SDW susceptibility exhibits Landau damping~\cite{Gerlach2016}. Whether this regime is characterized by a breakdown of Fermi liquid behavior, as observed in many unconventional superconductors, remains to be seen. 

{\em Acknowledgments.--}
The numerical simulations were performed on the CHEOPS cluster at RRZK
Cologne, the JUROPA/JURECA clusters at the Forschungszentrum J\"ulich,
and the ATLAS cluster at the Weizmann Institute. Y. S. and E. B. were
supported by the Israel Science Foundation under grant 1291/12, by the
US-Israel BSF under grant 2014209, and by a Marie Curie reintegration
grant. E. B. was supported by an Alon fellowship.  M. G. thanks the
Bonn-Cologne Graduate School of Physics and Astronomy (BCGS) for
support.

{\em Note added.--} While we were preparing this manuscript, Ref.~\cite{Wang2015} appeared, where a closely related model with O$(3)$ symmetry was studied. Our results are qualitatively similar to those of Ref.~\cite{Wang2015} where they overlap. 


\bibliography{AFM}


\appendix

\section{Details on the Monte Carlo simulations}
\label{sec:monte-carlo}

\subsection{Determinantal quantum Monte Carlo setup}
\label{sec:dqmc-setup}

The action~\eqref{eq:action} defines the partition function
\begin{align}
  \label{eq:1}
  Z = \int \differentialD(\vec{\varphi},\bar{\psi},\psi)\, e^{-S_{\varphi}-S_F}
  = \int \differentialD\vec{\varphi}\, e^{-S_{\varphi}} \Tr_{\psi}\left[ e^{-S_F}
  \right],
\end{align}
which we now bring into a form amenable to standard determinantal
quantum Monte Carlo (DQMC) methods \cite{Blankenbecler1981,White1989}
as they are presented in several pedagogical texts
\cite{Assaad2008,Assaad2002,DosSantos2003,LohJr.1992}.  We also
describe which measures need to be taken to attain a computational
time complexity no worse than the optimal $O(\beta \mathcal{N}^3)$,
where $\beta = 1/T$ is the inverse temperature and $\mathcal{N} = L^2$
the number of lattice sites.

To allow for an efficient numerical evaluation of the trace in
fermionic Fock space remaining in Eq.~\eqref{eq:1}, we discretize
imaginary time $\tau = \ell \dtau$, $\beta = m \dtau$ ($\dtau = 0.1$),
and after a symmetric Suzuki-Trotter decomposition we obtain
\begin{align}
  \label{eq:2}
  Z = \int \differentialD\vec{\varphi}\, e^{-\dtau \sum_{\ell = 1}^m
  L_{\varphi}(\ell \dtau)} \Tr_{\psi}\left[ \displaystyle \prod_{\ell
  = 1}^m \hat{B}_{\ell} \right] + O(\dtau^2) .
\end{align}
Here the operators $\hat{B}_{\ell}$ are given by
\begin{align}
  \label{eq:3}
  \hat{B}_{\ell}  = e^{-\frac{1}{2} \dtau \psi^{\dagger} K \psi}
  e^{-\dtau \psi^{\dagger} V_{\ell} \psi}
  e^{-\frac{1}{2} \dtau \psi^{\dagger} K \psi},  
\end{align}
with non-commuting matrices $K$ and $V_{\ell}$ and vectors of
fermionic operators
\begin{align}
  \label{eq:4}
  \psi^{\dagger} = \big(\psi^{\dagger}_{\alpha i \sigma}\big)
  ={}&\Big(
       \psi^{\dagger}_{x 1 \uparrow},   \dotsc, \psi^{\dagger}_{x \mathcal{N} \uparrow},
       \psi^{\dagger}_{y 1 \downarrow}, \dotsc, \psi^{\dagger}_{y \mathcal{N} \downarrow},
       \\
  {}&\phantom{\Big(}
      \psi^{\dagger}_{x 1 \downarrow}, \dotsc, \psi^{\dagger}_{x \mathcal{N} \downarrow},
      \psi^{\dagger}_{y 1 \uparrow},   \dotsc, \psi^{\dagger}_{y \mathcal{N} \uparrow},  
      \Big). \notag 
\end{align}
Explicitly, $K$ and $V_{\ell}$ are given by
\begin{align}
  \label{eq:5}
  K_{ij,\alpha\alpha',ss'}
  &=
    \delta_{ss'}\delta_{\alpha\alpha'}
    (-t_{\alpha,s,ij} - \mu \delta_{ij}),
    \notag\\
  V_{\ell;ij,\alpha\alpha',ss'}
  &=
    \lambda [\sigma_1]_{\alpha\alpha'}
    \delta_{ij} [\vec{s}\cdot\vec{\varphi}_i(\ell)]_{ss'}.
\end{align}
In this equation the Pauli matrix $\sigma_1$ acts on flavor indices,
while the Pauli matrices $\vec{s}$ act on spin indices.  We allow the
hopping constants $t$ to depend on spin in order to implement a
generalized magnetic field as described in
Sec.~\ref{sec:finite-size-effects} below.  In this $O(2)$-symmetric
model we have $\vec{\varphi}=(\varphi^1, \varphi^2)$.  Carrying out
the trace in Eq.~\eqref{eq:2} yields
\begin{align}
\label{eq:6}
  \Tr_{\psi}\left[ \displaystyle \prod_{\ell
  = 1}^m \hat{B}_{\ell} \right] =
  \det\left[ \Id + \displaystyle \prod_{\ell
  = 1}^m B_{\ell} \right]
  = \det G_{\varphi}^{-1}
\end{align}
with
$B_{\ell} = e^{-\frac{1}{2} \dtau K} e^{-\dtau V_{\ell}}
e^{-\frac{1}{2} \dtau K}$ \cite{Assaad2008}.  The matrix $G_{\varphi}$
is the equal-time Green's function evaluated for one bosonic spin
configuration $\{\vec{\varphi}_i(\ell \dtau)\}$.  After partitioning
the matrix exponentials into $\mathcal{N} \times \mathcal{N}$-sized
blocks, they read
\begin{align}
  \label{eq:7}
  e^{-\frac{\dtau}{2} K}
    &= \diag
      \left(
      e^{-\frac{\dtau}{2} K_{x}^{\uparrow}},
      e^{-\frac{\dtau}{2} K_{y}^{\downarrow}},
      e^{-\frac{\dtau}{2} K_{x}^{\downarrow}},
      e^{-\frac{\dtau}{2} K_{y}^{\uparrow}}
      \right),
      \notag\\
  e^{-\dtau V(\ell)}
    &=
      \begin{pmatrix}
        C & S & & \\
        S^{*} & C & & \\
        & & C & S^{*} \\
        & & S & C
      \end{pmatrix} 
     =
      \begin{pmatrix}
        \widetilde{V}(\ell) & \\
        & \widetilde{V}^{*}(\ell)
      \end{pmatrix}
\end{align}
with submatrices
\begin{align}
  \label{eq:8}
  C_{ij} &= \delta_{ij} \cosh \left( \dtau |\vec{\varphi}_j(\ell)|
           \right), \\
  S_{ij} &= \delta_{ij}\!\left[i \varphi^2_j(\ell) -
           \varphi^1_j(\ell)\right] \sinh \left(
           \dtau |\vec{\varphi}_j(\ell)| \right)
           \big/ |\vec{\varphi}_j(\ell)| . \notag
\end{align}
Under the condition
\begin{align}
  \label{eq:9}
  K_x^{\uparrow} &= K_x^{\downarrow, *}
  &                 
  {}&\mathrm{and}
  &
  K_y^{\downarrow} &= K_y^{\uparrow, *}
\end{align}
the Green's function decomposes into two blocks of size $2\mathcal{N}
\times 2\mathcal{N}$:
\begin{align}
  \label{eq:10}
  G_{\varphi} =
  \begin{pmatrix}
    \widetilde{G}_{\varphi} & \\
    & \widetilde{G}_{\varphi}^{*}
  \end{pmatrix}.
\end{align}
Hence we can write the partition function as
\begin{align}
  \label{eq:11}
  Z = \int\differentialD\vec{\varphi}\, e^{-\dtau \sum_{\ell = 1}^m
  L_{\varphi}(\ell \dtau)} \left| \det \widetilde{G}_{\varphi}^{-1}
  \right|^2 + O(\dtau^2),
\end{align}
which now is in a form that can be evaluated by Monte Carlo sampling
over space-time configurations
$\left\{ \vec{\varphi}_i(\ell) \right\}$.  Note that the probability
measure under the field integral is positive definite, which allows
for efficient sign-problem-free Monte Carlo simulations.  The
$O(2)$-symmetry allows us to restrict all fermionic evaluations to the
$(x\!\uparrow, y\!\downarrow)$-sector, which speeds up the most
expensive computations by a factor of 8 in comparison to the
$O(3)$-model.  From the matrix $G_{\varphi}$ we can compute arbitrary
fermionic equal-time observables via Wick's theorem and also access
imaginary-time-displaced correlation functions after the application
of matrices $B_{\ell}$ and $B_{\ell}^{-1}$. 

Generally, in the DQMC algorithm we frequently need to compute
products of the matrices $B_{\ell}$.  While the exponentials of
$V_{\ell}$ are sparse matrices and consequently their multiplication
has a computational cost of $O(\mathcal{\mathcal{N}}^2)$ only, even
for electron hopping restricted to nearest-neighbor sites, the
exponentials of the kinetic matrices $K_{\alpha}^s$ are densely
filled, which raises the cost of a single multiplication to
$O(\mathcal{\mathcal{N}}^3)$.  We avoid paying this cost by performing
a ``checkerboard'' decomposition \cite{Assaad2002}, where we divide
the whole set of lattice bonds into two groups, so that
$K_{\alpha}^{s\,(1,2)}$ are sums of commuting four-site hopping
matrices and
$K_{\alpha}^s = K_{\alpha}^{s\,(1)} + K_{\alpha}^{s\,(2)}$.  Applying
this decomposition for all $\alpha$ and $s$, we find
\begin{align}
  \label{eq:12}
  B_{\ell} &= e^{-\dtau K / 2} e^{-\dtau V_\ell} e^{-\dtau K / 2} \\
  &\approx
  e^{-\dtau K^{(1)} / 2} e^{-\dtau K^{(2)} / 2} e^{-\dtau V_\ell}
  e^{-\dtau K^{(2)} / 2} e^{-\dtau K^{(1)} / 2} \notag
\end{align}
and do not introduce any error of higher order than that already
present from the Suzuki-Trotter decomposition, yet save one power of
$\mathcal{N}$ in computational effort.

\subsection{Controlling finite-size effects}
\label{sec:finite-size-effects}

Simulations of metallic systems at low temperatures are particularly
susceptible to strong finite-size effects.  Since our numerical
methods limit us to the study of finite lattices, reducing the severity
of these effects is very important.  A dramatic reduction of
finite-size effects can be obtained in the presence of a perpendicular
magnetic field~\cite{Assaad2002a}.  In our simulations we thread a
single magnetic flux quantum $\Phi_0$ through the system, making sure
not to break condition~\eqref{eq:9} in order not to re-introduce a
sign problem.  Specifically, we add Peierls phase factors to the
hopping terms of the kinetic operator:
\begin{align}
  \label{eq:28}
  t_{\alpha,s,ij} \psi^{\dagger}_{x,i,s} \psi_{x,j,s}
  &\to e^{i A^{\alpha s}_{ij}} t_{\alpha,s,ij} \psi^{\dagger}_{\alpha,i,s}
  \psi_{\alpha,j,s}
  \quad \text{with} \notag \\
  A^{\alpha s}_{ij} &= \frac{2\pi}{\Phi_0}
  \int_{\mathbf{r}_i}^{\mathbf{r}_j} \differentiald \mathbf{x}
  \cdot \mathbf{A}^{\alpha s}
\end{align}
and choose the Landau gauge
$\mathbf{A}^{\alpha s}(\mathbf{x}) = -B^{\alpha s} x_2
\mathbf{\hat{e}}_1$.  The sign of the magnetic field depends on flavor
and spin indices $\alpha$, $s$ and its magnitude is the smallest
possible on the periodic $L \times L$ lattice:
\begin{align}
  \label{eq:31}
  B^{x\uparrow} = B^{y\downarrow} = -B^{x\downarrow} = -B^{y\uparrow}
  = \frac{\Phi_0}{L^2}.
\end{align}
Note that as $L \to \infty$ the original hopping constants are
restored.  To maintain translational invariance in presence of the
magnetic flux we impose special boundary conditions in the
$\mathbf{\hat{e}}_2$-direction
\begin{align}
  \label{eq:34}
  \psi_{\alpha, \mathbf{r} + L \mathbf{\hat{e}}_2, s} =
  \psi_{\alpha, \mathbf{r}, s} \exp\left(\frac{2\pi i}{\Phi_{0}} B^{\alpha s}
  L\, r_{1}\right),
\end{align}
while we retain regular periodic boundary conditions in
$\mathbf{\hat{e}}_1$-direction.  Explicitly, for nearest-neighbor
hopping, the phases read
\begin{align}
  \label{eq:35}
  A_{ij} = 
  \begin{cases}
    -\frac{2\pi}{\Phi_0} B^{\alpha s}\,i_2 &\text{ if } i_1 = 0,\dotsc,L-2
    \text{ and } j_1 = i_1 + 1 \\
    &\text{ or } i_1 = L-1 \text{ and } j_1 = 0, \\
    +\frac{2\pi}{\Phi_0} B^{\alpha s}\,i_2 &\text{ if } i_1 = 1,\dotsc,L-1
    \text{ and } j_1 = i_1 - 1 \\ 
    &\text{ or } i_1 = 0 \text{ and } j_1 = L-1, \\
    +\frac{2\pi}{\Phi_0} B^{\alpha s} L\, i_1 &\text{ if } i_2 = L-1
    \text{ and } j_2 = 0, \\
    -\frac{2\pi}{\Phi_0} B^{\alpha s} L\, i_1 &\text{ if } i_2 = 0
    \text{ and } j_2 = L-1, \Tstrut \\
    0 &\text{ otherwise,}
  \end{cases}
\end{align}
where the lattice site vectors are $\mathbf{r}_i = (i_1, i_2)$ and
$\mathbf{r}_j = (j_1, j_2)$, which we index from $0$ to $L-1$ in each
direction.

\subsection{Local and global updates}
\label{sec:local-global-update}

The foundation of our Monte Carlo simulations of the lattice field
theory~\eqref{eq:11} is the Metropolis algorithm, where a proposed
change of a bosonic field configuration
$\left\{ \vec{\varphi} \right\} \to \left\{ \vec{\varphi}\,{}'
\right\}$ is accepted with probability
\begin{align}
  \label{eq:13}
  p = \min \left\{ 1, e^{-(S_{\varphi}' - S_{\varphi})} \left|
  \frac{\det \widetilde{G}_{\varphi}}
  {\det \widetilde{G}_{\varphi}'} \right|^2 \right\} .
\end{align}
The most important type of proposed changes is the local update.  For
one sweep of local updates we attempt to change the orientations and
lengths of individual bosonic spins
$\vec{\varphi}_i(\tau = \ell \dtau)$ chosen sequentially from the
space-time lattice.  During an initial equilibration phase we
generally tune the size of the box from which the new spin is chosen
such that about $50\%$ of all local updates are accepted.  For local
updates the determinant ratio in \eqref{eq:13} is given by
\begin{align}
  \label{eq:14}
  \det \widetilde{G}_{\varphi} / \det \widetilde{G}_{\varphi}'
  &= \det[ \Id + \Delta(\Id - \widetilde{G}_{\varphi}(\ell)) ],
\end{align}
where
$\widetilde{G}_{\varphi}(\ell) \equiv [ \Id + \widetilde{B}_{\ell}
\dotsm \widetilde{B}_1 \widetilde{B}_m \dotsm \widetilde{B}_{\ell+1}
]^{-1}$ and
$\Delta \equiv \widetilde{B}'_{\ell} \widetilde{B}^{-1}_{\ell} - \Id =
e^{-\dtau \widetilde{V}'_{\ell}} e^{\dtau \widetilde{V}_{\ell}} -
\Id$.  We find that an expansion by minors reduces the determinant in
Eq.~\eqref{eq:14} to that of a $2 \times 2$-matrix
$M = \Id_2 + (\Id_2 - \widetilde{G}^i) \cdot \Delta^i$, where
$\Delta^i = \Delta[i:\,:\mathcal{N}, i:\,:\mathcal{N}]$,
$\widetilde{G}^i = \widetilde{G}_{\varphi}(\ell)[i:\,:\mathcal{N},
i:\,:\mathcal{N}]$, and the slice index notation corresponds to the
four sole non-zero entries of $\Delta$.  Thus the acceptance
probability can be computed in constant time.

After an accepted local update the Green's function matrix must be
updated:
\begin{align}
  \label{eq:16}
  \widetilde{G}'_{\varphi}(\ell) = \widetilde{G}_{\varphi}(\ell) 
  [ \Id + \Delta(\Id - \widetilde{G}_{\varphi}(\ell)) ]^{-1},
\end{align}
where we can again exploit the sparseness of $\Delta$ and replace the
inversion of the $2\mathcal{N} \times 2\mathcal{N}$-matrix by that of
a $2 \times 2$-matrix if we make use of the Sherman-Morrison-Woodbury
formula~\cite{Golub2013}.  In this manner we find
\begin{align}
  \label{eq:17}
  \widetilde{G}'_{\varphi}(\ell) = \widetilde{G}_{\varphi}(\ell) +
  {}&(\widetilde{G}_{\varphi}(\ell)[:, i:\,:\mathcal{N}] \cdot \Delta^i) \\
  {}&\cdot \left( M^{-1} \cdot \left\{(G_{\varphi}(\ell) -
      \Id)[i:\,:\mathcal{N}, :]\right\}\right)
      \notag ,
\end{align}
which, if the matrix products are carried out in the order indicated
by the parentheses, can be evaluated with only $O(\mathcal{N}^2)$
operations.  In consequence, one total sweep of local updates has a
time complexity of $O(\beta \mathcal{N}^3)$.

Using only these local updates, Monte Carlo simulations of this model
require relatively long thermalization periods without measurements to
equilibrate the system and then generally show long statistical
autocorrelation times, which are amplified near the magnetic phase
transition by the effect of critical slowing down.  To counteract
these effects we adopt two tools: a simple global update and a replica
exchange mechanism.

The global move consists of adding a constant random displacement
$\vec{\delta}$ to all bosonic spins:
$\vec{\varphi}_i(\tau) \to \vec{\varphi}_i(\tau) + \vec{\delta}$.  To
evaluate the acceptance probability~\eqref{eq:13} we compute
$\widetilde{G}_{\varphi}'$ and its determinant from scratch, which takes
$O(\beta \mathcal{N}^3)$ operations.  At times we found it also
helpful to combine this move with the Wolff single cluster
algorithm~\cite{Wolff1989}.  Here we ignore the fermionic part of the
action while we construct and flip a cluster of spins, then we add the
global displacement, and finally we decide on accepting the joint move
according to Eq.~\eqref{eq:13}.

\subsection{Replica exchange}
\label{sec:replica-exchange}

For the replica exchange or parallel tempering
scheme~\cite{Geyer1991,HukushimaNemoto} we consider an extended
ensemble composed of multiple grand-canonical ensembles with the
parameter $r$ in $S_{\varphi}$ taking on different values
$r_1 < r_2 < \ldots < r_K$ such that the partition function is given
by a product $\mathcal{Z} = \prod_{\kappa = 1}^K Z(r_{\kappa})$,
\begin{align}
  \label{eq:15}
  \mathcal{Z} = 
  \int \differentialD (\vec{\varphi}_{1}, \ldots, \vec{\varphi}_{K})
  \prod_{\kappa = 1}^K    
  e^{-S_{\varphi}(r_{\kappa}, \{ \vec{\varphi}_{\kappa} \})}
  \left| \det \widetilde{G}_{\varphi_{\kappa}}^{-1} \right|^2,
\end{align}
where $\widetilde{G}_{\varphi_{\kappa}}$ does not depend on
$r_{\kappa}$.  In the Monte Carlo simulation we then have in parallel
a separate replica of the system for every $r_{\kappa}$, each being
represented by a different system configuration.  The control
parameter $r$ is treated as a dynamical variable by allowing exchanges
of the configurations between replicas of different parameter values.
In this way shorter autocorrelation times at high $r$ can be utilized
to accelerate the simulation across the phase transition and in the
low-$r$ region.  To achieve this we need to construct a Monte Carlo
move between replicas, which will supplement the single-replica local
and global updates that are still carried out as in regular canonical
simulations.  In such an update we propose the exchange of
configurations $\left\{ \vec{\varphi} \right\}$ and
$\left\{ \vec{\varphi}\,{}' \right\}$ between the $\kappa$-th and
$\eta$-th replicas.  To ensure detailed balance we require
\begin{multline}
  \label{eq:19}
  P(\dotsc, \vec{\varphi}, r_{\kappa}, \dotsc, \vec{\varphi}\,{}', r_{\eta},
  \dotsc) W(\vec{\varphi}, r_{\kappa} | \vec{\varphi}\,{}', r_{\eta}) \\
  = P(\dotsc, \vec{\varphi}\,{}', r_{\kappa}, \dotsc, \vec{\varphi}, r_{\eta},
  \dotsc) W(\vec{\varphi}\,{}', r_{\kappa} | \vec{\varphi}, r_{\eta}),
\end{multline}
where $P(\vec{\varphi}_1, r_1, \ldots, \vec{\varphi}_K, r_K)$ is the
equilibrium probability of a set of system configurations
$\left\{ \vec{\varphi}_{\kappa} \right\}$ associated to parameters
$r_{\kappa}$ in the extended ensemble and $W$ is the transition
probability for a replica configuration exchange.  The ratio of these
transition probabilities is
\begin{align}
  \label{eq:20}
  \frac{W(\vec{\varphi}, r_{\kappa} | \vec{\varphi}\,{}', r_{\eta})}
  {W(\vec{\varphi}\,{}', r_{\kappa} | \vec{\varphi}, r_{\eta})}
  = \frac{
  e^{-S_{\varphi}(r_{\kappa},\{ \vec{\varphi}\,{}' \})
  -S_{\varphi}(r_{\eta}, \{ \vec{\varphi} \})
  }}{
  e^{-S_{\varphi}(r_{\kappa}, \{ \vec{\varphi} \})
  -S_{\varphi}(r_{\eta}, \{ \vec{\varphi}\,{}' \})
  }}
  = e^{-\Delta},
\end{align}
where
$\Delta = (r_{\kappa} - r_{\eta}) \cdot \frac{1}{2} \int_0^{\beta}d
\tau \sum_i \left[ \vec{\varphi}\,{}'_i(\tau)^2 -
  \vec{\varphi}_i(\tau)^{2} \right]$ and we note that the fermion
determinants have canceled.  To fulfill the relation~\eqref{eq:20} we
choose exchange probabilities according to the Metropolis criterion
\begin{align}
  \label{eq:21}
  W(\vec{\varphi}, r_{\kappa} | \vec{\varphi}\,{}', r_{\eta}) = \min
  \left\{ 1, e^{-\Delta} \right\}.
\end{align}
In our simulations we only propose exchanges between adjacent pairs of
control parameter values.  At high temperatures we achieve good
diffusion with a simple linear spacing of the values of $r$.  At lower
temperatures, however, the magnetic phase transition constitutes a
more significant barrier to the random walk in $r$-space.  Here we
have used a feedback-optimized distribution of
$r$-values~\cite{Katzgraber2006,Trebst2006}, which effectively
clusters the $r_{\kappa}$ around $r_{\text{SDW}}(T)$, easing diffusion
and significantly lowering autocorrelation times.  Since the exchange
algorithm following Eq.~\eqref{eq:20} does not require the
recomputation of Green's functions or the evaluation of their
determinants, it poses very little overhead in computation or
communication.  This allows us to perform a replica-exchange sweep
after every single sweep of canonical updates, which has been very
beneficial for obtaining sufficient statistics to resolve the magnetic
phase diagram.

\subsection{Time series reweighting}
\label{sec:reweighting}

The structure of the action~\eqref{eq:action}, where the
$r$-dependence is fully contained in the bosonic part $S_{\varphi}$,
allows to easily relate the canonical probability distribution of a
configuration $\{ \vec{\varphi} \}$ at a tuning parameter value $r$,
$p_r(\vec{\varphi} )$, to the distribution at another value $r'$:
$p_{r'}( \vec{\varphi}) \propto e^{-(r'-r) E(\vec{\varphi})} p_r(
\vec{\varphi})$, where
$E(\vec{\varphi})=\frac{1}{2}\int_0^{\beta}\differentiald \tau \sum_i
\vec{\varphi}_i(\tau)^2$.  From this relation one finds an
expression for the expectation value of an observable $\mathcal{O}$ at
$r'$ in terms of expectation values at $r$, which in turn can be
estimated by time series averages from a Monte Carlo simulation
carried out at $r$:
\begin{align}
  \label{eq:22}
  \left\langle \mathcal{O} \right\rangle_{r'} =
  \frac{ \langle \mathcal{O} e^{-(r'-r)E} \rangle_r }
  { \langle e^{-(r'-r)E} \rangle_r  } \approx
  \frac{ \sum_n \mathcal{O}_n e^{-(r'-r)E_n}  }
  { \sum_n e^{-(r'-r)E_n} },
\end{align}
where $n$ goes over the series of measured samples and $\mathcal{O}_n$
and $E_n$ are computed from the same system configuration.  This
reweighting procedure~\cite{Ferrenberg1988} is effective over quite a
wide range around $r$.

From our replica exchange simulations we have Monte Carlo data for
multiple close values of $r$.  We can use the combined information
from these time series for $r_{\kappa}$, $\kappa = 1, \dotsc, K$, to
obtain improved observable estimates at $r_1 \le r \le r_K$ by
multiple histogram reweighting~\cite{Ferrenberg1989,Chodera2007}.  To
do so we write the expectation value as
\begin{align}
  \label{eq:23}
  \left\langle \mathcal{O} \right\rangle_{r} &= \frac{
  \int \differentiald E \Omega(E) e^{-r E} \mathcal{O}(E)
  }{
  \int \differentiald E \Omega(E) e^{-r E}  
  } \quad \mathrm{with} \notag\\
  \mathcal{O}(E) &= \frac{
  \int \differentialD \vec{\varphi}
  \delta(E[\vec{\varphi}] - E) \mathcal{O}[\vec{\varphi}]
  }{
  \int \differentialD \vec{\varphi} \delta(E[\vec{\varphi}] - E) 
  },
\end{align}
where all non-$r$-dependent parts of the action are contained in the
density of states $\Omega(E)$.  We discretize $E$ into levels
$E_{\alpha}$ spaced $\Delta E$ apart and search the optimal estimator
for $\Omega(E_{\alpha})$, which reads
\begin{align}
  \label{eq:24}
  \hat{\Omega}_{\alpha} = \frac{ \sum_{\kappa} H_{\alpha\kappa}
  [ g_{\alpha\kappa} (1 - \Delta E \hat{\Omega}_{\alpha}
  e^{-r_{\kappa} E_{\alpha} + f_{\kappa}}) ]^{-1}}
  { \sum_{\kappa} M_{\kappa} \Delta E e^{-r_{\kappa} E_{\alpha} +
  f_{\kappa}} [ g_{\alpha\kappa} (1 - \Delta E \hat{\Omega}_{\alpha}
  e^{-r_{\kappa} E_{\alpha} + f_{\kappa}}) ]^{-1} }.
\end{align}
Here $H_{\alpha\kappa}$ is the count of samples with
$E \in [E_{\alpha}, E_{\alpha} + \Delta E)$ in the time series with
$r=r_{\kappa}$, $g_{\alpha\kappa}$ is a statistical inefficiency
factor related to the integrated autocorrelation time of the indicator
function for this count, $M_{\kappa}$ is the total number of
samples for $r_{\kappa}$ and $f_{\kappa} = -\ln Z(r_{\kappa})$ is
given by
\begin{align}
  \label{eq:25}
  f_{\kappa} = -\ln \sum_{\alpha} \hat{\Omega}_{\alpha} \Delta E
  e^{-r_{\kappa} E_{\alpha}}.
\end{align}
Empirically, we find it adequate to set $g_{\alpha\kappa} \equiv 1$.
Iteration of Eqs.~\eqref{eq:24} and \eqref{eq:25} yields a converged
estimate of $\hat{\Omega}_{\alpha}$ and following Eq.~\eqref{eq:23} we
compute the estimate of $\langle O \rangle(r)$ as a weighted average
of the time series of $\mathcal{O}$ for the different $r_{\kappa}$:
\begin{align}
  \label{eq:26}
  \hat{\mathcal{O}}(r) = \frac{ \sum_{\kappa=1}^{K} \sum_{n=1}^{M_{\kappa}}
  \mathcal{O}_{\kappa n} w_{\kappa n}(r)}
  { \sum_{\kappa=1}^{K} \sum_{n=1}^{M_{\kappa}} w_{\kappa n}(r) }
\end{align}
with weights
\begin{align}
  \label{eq:27}
  w_{\kappa n}(r) = \sum_{\alpha} \frac{\psi_{\alpha \kappa n}
  \hat{\Omega}_{\alpha} e^{-r E_{\alpha}}}{\sum_{\kappa} H_{\alpha\kappa}},
\end{align}
where $\psi_{\alpha \kappa n}$ is the indicator function for
$E \in [E_{\alpha}, E_{\alpha} + \Delta E)$ evaluated at the $n$-th
sample of the time series for $r_{\kappa}$.

The multiple histogram reweighting method allows us to finely
interpolate between the original values $r_{\kappa}$ of our
simulations.  In addition it provides a reduction of statistical error
bars in the reweighted estimates compared to averages from single time
series.  In this work we have used the method for bosonic observables
related to the magnetic transition, although it can easily be extended
to all fermionic observables.


\section{Magnetic transition}
\label{sec:magnetic_transition}

In the thermodynamic limit the model described by \eqref{eq:action}
cannot show magnetic long-range order at any $T>0$ as stated by the
Mermin-Wagner theorem~\cite{Mermin1966}.  Nevertheless, a
finite-temperature phase transition of the
Berezinsky-Kosterlitz-Thouless (BKT)
type~\cite{Berezinskii1971,Kosterlitz1973,Kosterlitz1974} is not
precluded in this $O(2)$-symmetric model.  Defining a local
magnetization
\begin{align}
  \label{eq:18}
  \vec{m}_i = \frac{1}{\beta} \int_0^{\beta} \differentiald \tau
  \vec{\varphi}_i(\tau),
\end{align}
in such a scenario the total magnetization density
$\vec{m} = \frac{1}{L^2} \sum_i \vec{m}_i$ vanishes in the
thermodynamic limit $L^2 \to \infty$ even below the transition
temperature $T_{\text{SDW}}$, where only finite systems will have
quasi-long-range order with
$\left\langle |\vec{m}| \right\rangle \ne 0$. At temperatures
approaching $T_{\text{SDW}}$ from above the correlation length $\xi$
diverges exponentially
\begin{alignat}{2}
  \label{eq:29}
  \xi &\sim \exp\left(b(T-T_{\text{SDW}})^{-\nu}\right),\quad & T &\to
  T_{\text{SDW}}^+,
\end{alignat}
with a critical exponent $\nu=1/2$ and it stays infinite for all
$T \le T_{\text{SDW}}$, so that the entire low-temperature phase is
critical.  Spatial correlation functions of the local magnetization
fluctuations decay exponentially above $T_{\text{SDW}}$ and with a
power law below $T_{\text{SDW}}$:
\begin{align}
  \label{eq:30}
  \left\langle
  \vec{m}_i \cdot \vec{m}_{i+\mathbf{x}}
  \right\rangle 
  \sim
  \begin{cases}
    e^{-|\mathbf{x}| / \xi}, & T > T_{\text{SDW}} , \\
    |\mathbf{x}|^{-\eta(T)}, & T \le T_{\text{SDW}} .
  \end{cases}
\end{align}
The critical exponent $\eta$ depends on temperature with
$\eta(T_{\text{SDW}}) = 1/4$.  Following e.g. Refs.~\cite{Cuccoli1995,Wysin2005}
we study the spin-density wave susceptibility
\begin{align}
  \label{eq:32}
  \chi =  \beta \sum_i \langle \vec{m}_i \cdot \vec{m}_0 \rangle
  = \int_0^{\beta} \!\differentiald \tau \sum_i \langle
  \vec{\varphi}_i(\tau) \vec{\varphi}_0(0) 
  \rangle = \beta L^2 \langle \vec{m}^2 \rangle
\end{align}
and from Eq.~\eqref{eq:30} expect a finite-size scaling behavior like
\begin{align}
  \label{eq:33}
  \chi \sim L^{2-\eta}
\end{align}
with $\eta > 0$ for $T \le T_{\text{SDW}}$ and slightly higher
temperatures, where $\xi$ still exceeds $L$.  We identify points
$(r,T)$ in the phase diagram where Eq.~\eqref{eq:33} can be fitted
well to our data with $\eta \le 1/4$ as belonging to the
quasi-long-range ordered SDW phase.

\begin{figure}[t]
  \includegraphics[width=\linewidth]{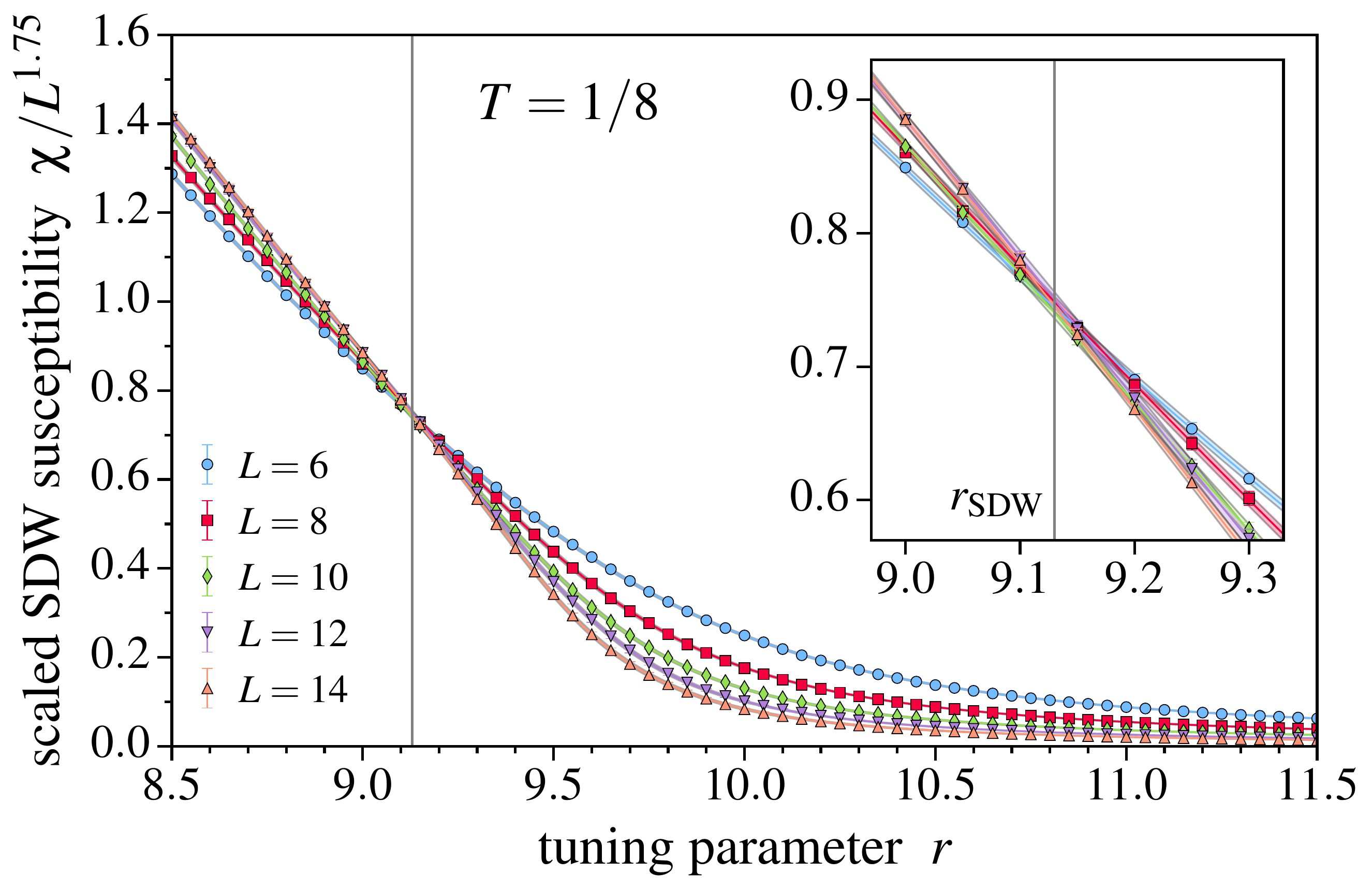}
  \caption{Scaled SDW susceptibility $\chi / L^{2-\eta_c}$ with
    $\eta_c = 1/4$ for $T=1/8$ and various system sizes.  Symbols with
    error bars are estimates from single-$r$ data.  Continuous lines
    with surrounding error regions are results of the
    multiple-histogram reweighting analysis.}
  \label{fig:chi-intersection}
\end{figure}

\begin{figure}
  \includegraphics[width=\linewidth]{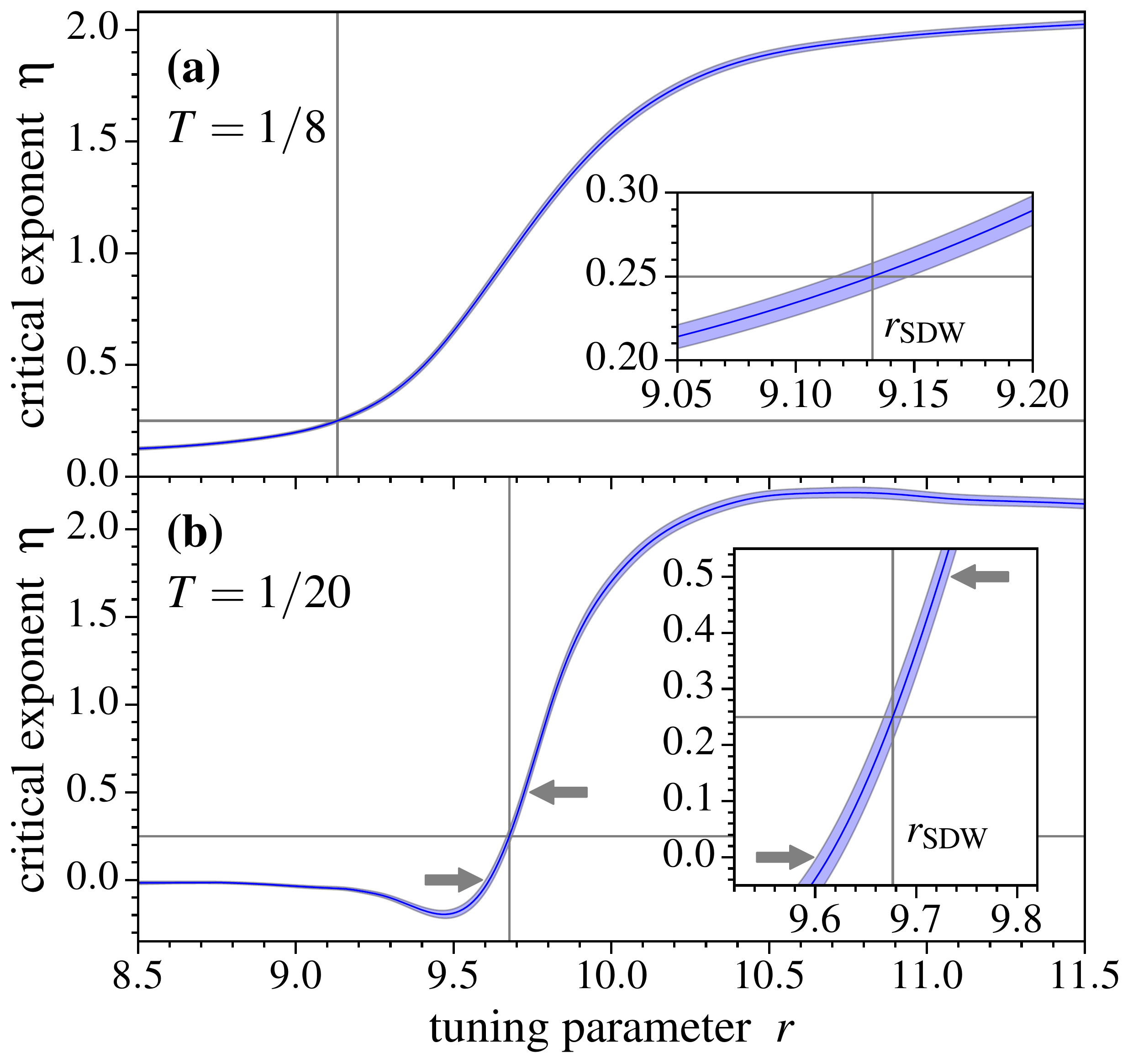}
  \caption{Critical correlation exponent $\eta(r)$ at (a)~$T=1/8$ and
    (b)~$T=1/20$ for estimation of $r_{\text{SDW}}$, where
    $\eta=1/4$. The continuous line shows the result of fitting
    $\ln \chi = \alpha + (2-\eta) \ln L$ to the DQMC data, which has
    been interpolated by reweighting.  The shaded surrounding region
    indicates the statistical error.  The fits in (b) are of low
    quality.  Here the arrows indicate a wider estimate of the error
    on $r_{\text{SDW}}$, allowing for $\eta \in [0, 0.5]$.}
  \label{fig:eta-r-beta8-20}
\end{figure}

\begin{figure}
  \centering
  \includegraphics[width=0.88\linewidth]{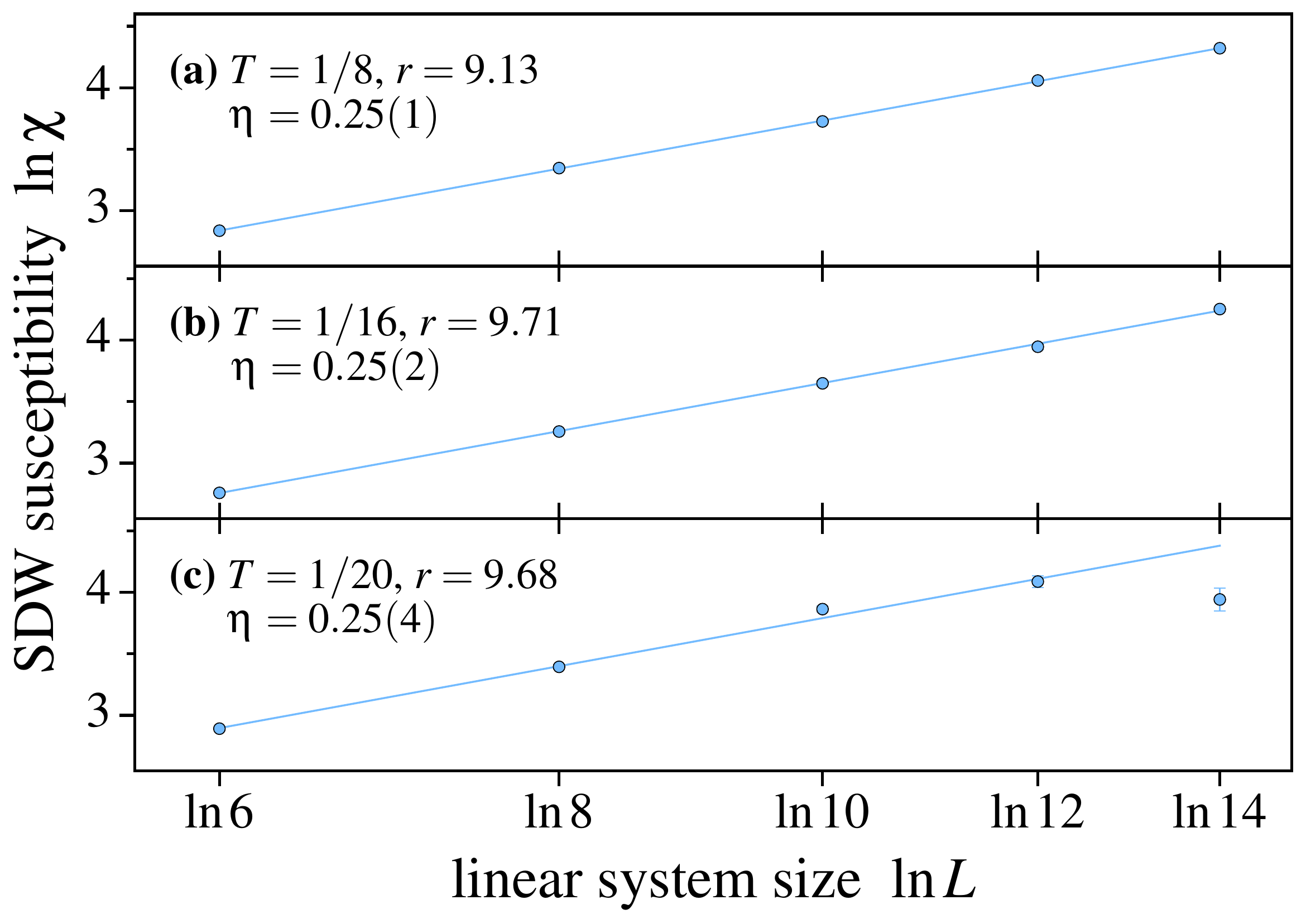}
  \caption{Best fits of $\ln \chi = \alpha + (2-\eta) \ln L$ with
    $\eta = 1/4$ for (a)~$T=1/8$, (b)~$T=1/16$, and (c)~$T=1/20$.}
  \label{fig:eta-fit-examples}
\end{figure}

At constant $T$ we scan over $r$ and fit the relation
$\ln \chi = \alpha + (2-\eta) \ln L$ to our data to determine
$\eta(r)$.  Then we search for $\eta(r_{\text{SDW}}) = 1/4$ to find
where $T=T_{\text{SDW}}$.  The reweighting technique described in
Sec.~\ref{sec:reweighting} provides us with high resolution in $r$ to
pinpoint $r_{\text{SDW}}$.  In Fig.~\ref{fig:chi-intersection} we show
that the intersection point of the scaled SDW susceptibility
$\chi / L^{2-\eta_c}$ with $\eta_c = 1/4$ coincides approximately with
this estimate for $r_{\text{SDW}}$.  Fig.~\ref{fig:eta-r-beta8-20}
illustrates the dependence of the estimated $\eta$ on $r$, while
Fig.~\ref{fig:eta-fit-examples} shows representative examples for fits
with $\eta=1/4$.  As it is apparent there, the scaling
relation~(\ref{eq:33}) fits our DQMC well for $T \gtrsim 1/16$, but
for $T \le 1/20$ we cannot find good agreement with the power law on
the range of lattice sizes we have accessed.  To account for a
systematic error at these low temperatures we give a wider estimate of
the error on $r_{\text{SDW}}$, allowing for values of
$\eta \in [0, 0.5]$ (see Fig.~\ref{fig:eta-r-beta8-20}b), while at
higher temperatures we provide purely statistical error estimates
computed from the variance-covariance matrix of the linear fit.  A
precise quantification of the systematic error in this finite-size
scaling analysis would require system sizes $L$ that are larger by
orders of magnitude and hence out of computational reach.  In
Table~\ref{tab:eta-fit-results} we summarize our results for
$r_{\text{SDW}}(T)$ as determined from fits over five values
$L = 6, \dotsc, 14$, which are also plotted in
Fig.~\ref{fig:phasediagram} in the main text, and show in comparison
results for a reduced range $L=8, \dotsc, 14$.  The data points for
$T_{\text{SDW}} \le 1/20$ in Fig.~\ref{fig:phasediagram}, where we
were not able to obtain a good fit, are connected by dotted lines.

\begin{table}
  \caption{Location of the SDW transition point $r_{\text{SDW}}$ for
    different temperatures $T$ as estimated by fitting $\ln \chi =
    \alpha + (2-\eta) \ln L$ and searching for $\eta = 1/4$ for two
    ranges of system sizes $L=6,\dotsc,14$ ($n=5$ data points) and 
    $L=8,\dotsc,14$ ($n=4$).  $\cdof = {\chi^2}/({n - 2})$ 
    is a measure to help with the estimation of the validity of the
    fit.  For $T \le 1/20$ the equation does not fit the data well and
    $\cdof$ is larger than unity. }
  \label{tab:eta-fit-results}
  \begin{ruledtabular}
    \begin{tabular}{
      r
      D{.}{.}{1.3}
      D{.}{.}{2.1}
      D{.}{.}{1.4}
      D{.}{.}{2.1}
      }
      & \multicolumn{2}{c}{$L=6,\dotsc,14$} & \multicolumn{2}{c}{$L=8,\dotsc,14$}
      \\
      \cline{2-3} \cline{4-5}
      \mc{$1/T$} & \mc{$r_{\text{SDW}}$ \Tstrut\Bstrut} & \mc{$\cdof$} & \mc{$r_{\text{SDW}}$} & \mc{$\cdof$}  \\
      \hline \Tstrut
      4  & 7.54(3) & 0.6  & 7.6(1)   & 0.3 \\
      5  & 8.10(3) & 1.4  & 8.07(5)  & 1.6 \\
      6  & 8.51(4) & 1.3  & 8.499(2) & 1.1 \\
      8  & 9.13(2) & 0.9  & 9.12(3)  & 1.4 \\
      10 & 9.53(1) & 0.4  & 9.52(3)  & 0.5 \\
      12 & 9.72(1) & 1.8  & 9.73(3)  & 2.5 \\
      13 & 9.73(1) & 0.1  & 9.73(1)  & 0.1 \\
      14 & 9.72(1) & 4.0  & 9.76(1)  & 0.3 \\
      16 & 9.71(1) & 0.5  & 9.71(1)  & 0.6 \\
      \cline{2-5}      
      \Tstrut
      20 & 9.68(8) & 10.2 & 9.7(1)  & 13.6\\
      26 & 9.68(5) & 11.0 & 9.66(7)  & 7.8 \\
      30 & 9.66(6) & 4.4  & 9.62(9)  & 3.7 \\
    \end{tabular}
  \end{ruledtabular}
\end{table}

\begin{figure}[th]
  \centering
  \includegraphics[width=\linewidth]{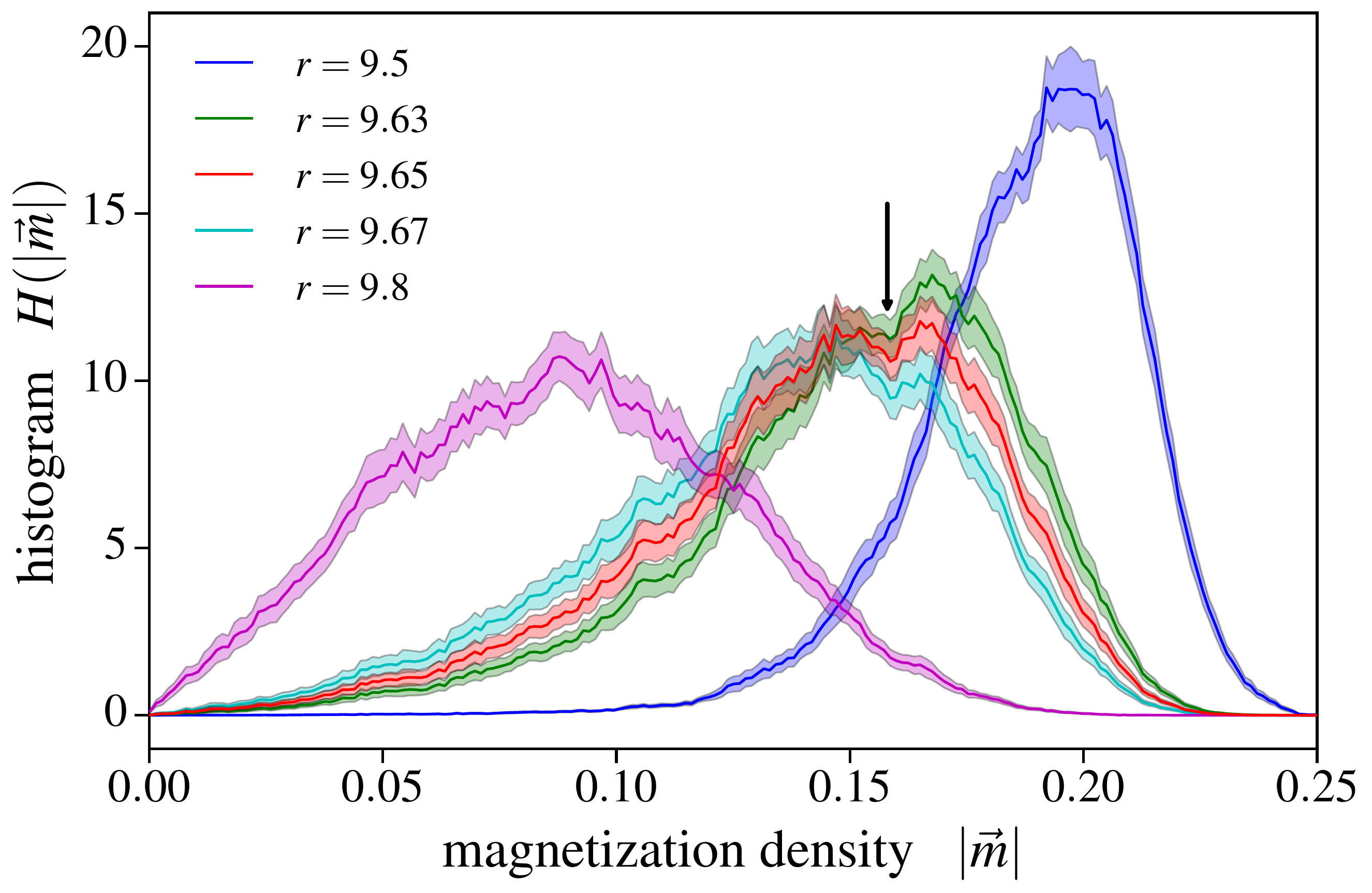}  
  \caption{Low-temperature histograms at $T=1/20$ and $L=14$ of the
    finite-system magnetization density $|\vec{m}|$ show a small
    suppression between two peaks at $r \approx 9.65$ close to the
    estimated location of the phase transition.  This may be a
    signature of a weak first-order transition.}
  \label{fig:histograms}
\end{figure}

The temperature below which the scaling law~(\ref{eq:33}) may be
invalid lies under the superconducting $T_c$.  There we have some
indications that the magnetic transition could be weakly first-order
and not of the BKT type.  In extensive simulations at $T=1/20$ for the
largest system size $L=14$ accessed by us the histograms of the
finite-system magnetization density show a shallow double-peak
structure when we tune $r$ to an intermediate value between the
magnetically quasi-long-range ordered and disordered phases, see
Fig.~\ref{fig:histograms}.  The location of this point is marked by a
cross in Fig.~\ref{fig:phasediagram} in the main text.  If this dip
grows deeper for larger systems, this bimodal distribution can be
understood as a sign of phase coexistence at a first-order
transition~\cite{JankeFirstOrder}.  In our DQMC simulations close to
the approximate transition point we also observe noticeably longer
statistical autocorrelation times at $T \le 1/20$ than at higher
temperatures, which may be explained by the first-order transition and
would also make it very cumbersome to obtain sufficient statistics to
resolve these histograms for larger $L$.

\FloatBarrier

\section{Diamagnetic response and the identification of the superconducting $T_c$}
\label{sec:diamagnetic}
In the appropriate gauge, the linear response of the system to a static, orbital magnetic field $B(\mathbf{q})$ is given by 
\begin{equation}
j_x(\mathbf{q}) = - 4 K_{xx}(\mathbf{q}) A_x(\mathbf{q}), 
\end{equation}
where $A_x(\mathbf{q})=iB(\mathbf{q})/q_y$ is the vector potential in an appropriate gauge, and 
\begin{equation}
K_{xx}(\mathbf{q}) \equiv \frac{1}{4} \left[\Lambda_{xx} (q_x\rightarrow 0, q_y=0)  -  \Lambda_{xx} (\mathbf{q})  \right].
\label{eq:Kxx}
\end{equation}
Here, $\Lambda_{xx}$ is the current-current correlator
\begin{equation}
\Lambda_{xx}(\mathbf{q}) = \sum_{i}\int_0^\beta d\tau e^{-i\mathbf{q} \cdot \mathbf{r}_i} \langle j_x(\mathbf{r}_i,\tau) j_x(0,0)\rangle,
\end{equation}
and the current density operator is given by $j_x(\mathbf{r}_i) = \sum_{\alpha, s} i t_{\alpha i s} \psi^\dagger_{\alpha i s} \psi_{\alpha i s} + \mathrm{H.c.}$,  where $\mathbf{r}_j = \mathbf{r}_i + \hat{x}$.

In the normal state, the magnetization is given by $-4 \mathrm{lim}_{q_y \rightarrow 0} K_{xx}/q_y^2$. We note in passing that for general lattice models, the magnetic response can be of either sign. For the band parameters chosen in the text, the response in the non-interacting ($\lambda=0$) case is paramagnetic.

\begin{figure}[th]
  \centering
  \includegraphics[width=\linewidth]{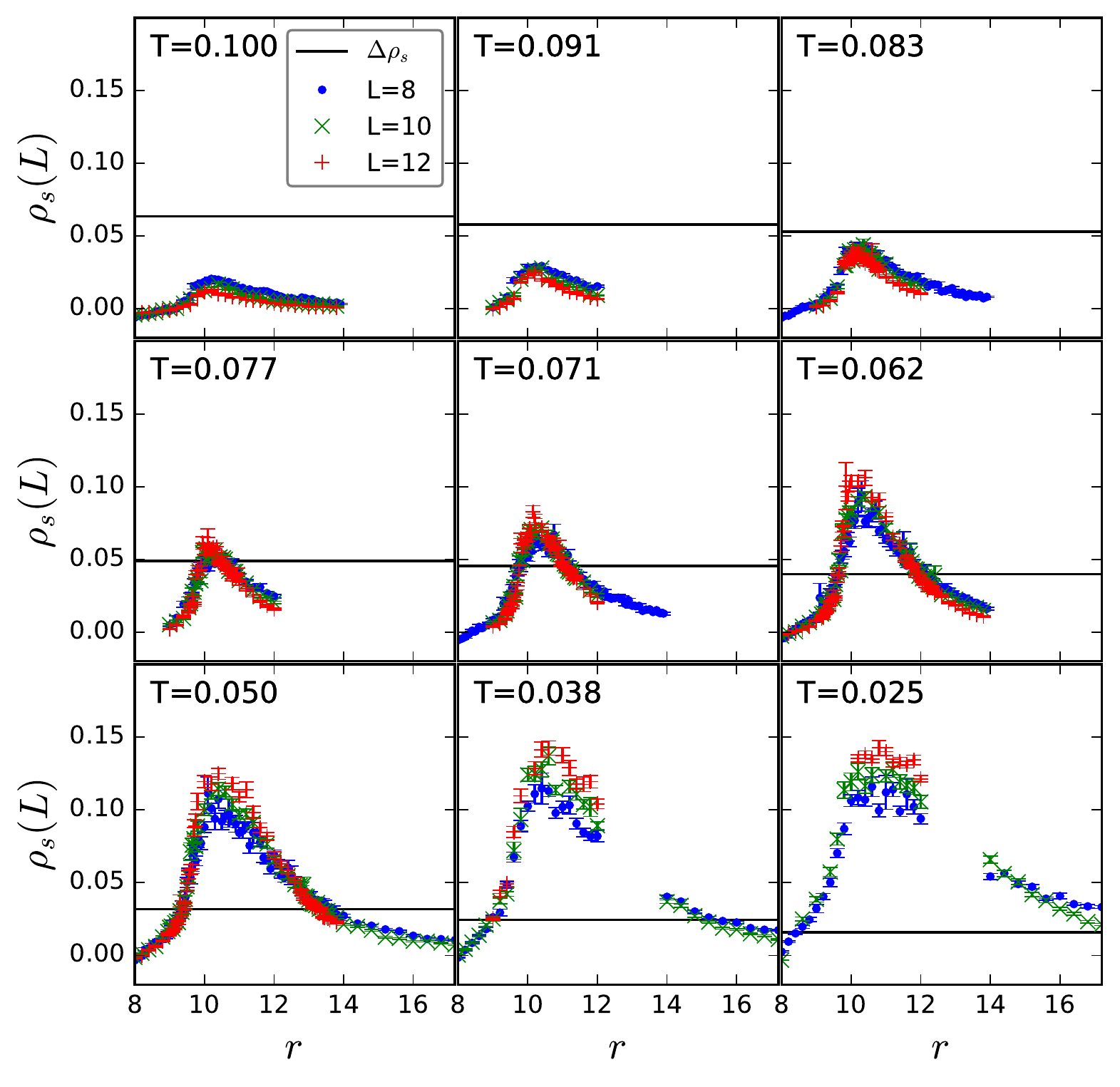}
  \caption{$\rho_s(L)$, as defined in the text, for system sizes $L=8,10,12$ across the phase diagram. The solid line indicates the universal value $\Delta \rho_s = \frac{2T}{\pi}$ expected at the BKT transition.}
  \label{fig:rho_s_vs_r}
\end{figure}

To identify the superconducting transition, we employ the analysis of Ref.~\cite{Paiva2004}. The superfluid density is given by~\cite{Scalapino1993}
\begin{equation}
\rho_s = \lim_{q_y\rightarrow 0} \lim_{L\rightarrow \infty} K_{xx}(q_x=0, q_y) 
\label{eq:rho_s}
\end{equation}
Here, for convenience, we will use the notation $\rho_s(L)= K_{xx}(q_x=0, q_y=2\pi/L)$, whose limit when $L\rightarrow \infty$ is the superfluid density. 
At the BKT transition, the superfluid density changes discontinuously by a universal amount, $\Delta \rho_s= \frac{2T}{\pi}$. Figure \ref{fig:rho_s_vs_r} shows $\rho_s(L)$ across the phase diagram for multiple temperatures. For each temperature we identify the values of $r$ at which $\rho_s(L)>\Delta \rho_s$ as the superconducting phase. The finite-size effects are not very substantial (except perhaps at large $r$ at the lowest temperature $T=0.025$), and are our main source of error in determining the superconducting phase boundary.

The analysis of the superfluid density does not rely on a particular ansatz for the superconducting order parameter. To determine the symmetry of the superconducting order parameter, we consider the uniform susceptibility $P_\eta(\mathbf{q}=0)$, as defined in~\eqref{eq:susc}. Close to the BKT transition, the susceptibility of the appropriate pairing channel scales as $L^{2-\eta}$, where $\eta$ varies continuously with temperature, reaching the value $\eta=0.25$ at $T_c$. 

\begin{figure}[th]
  \centering
  \includegraphics[width=\linewidth]{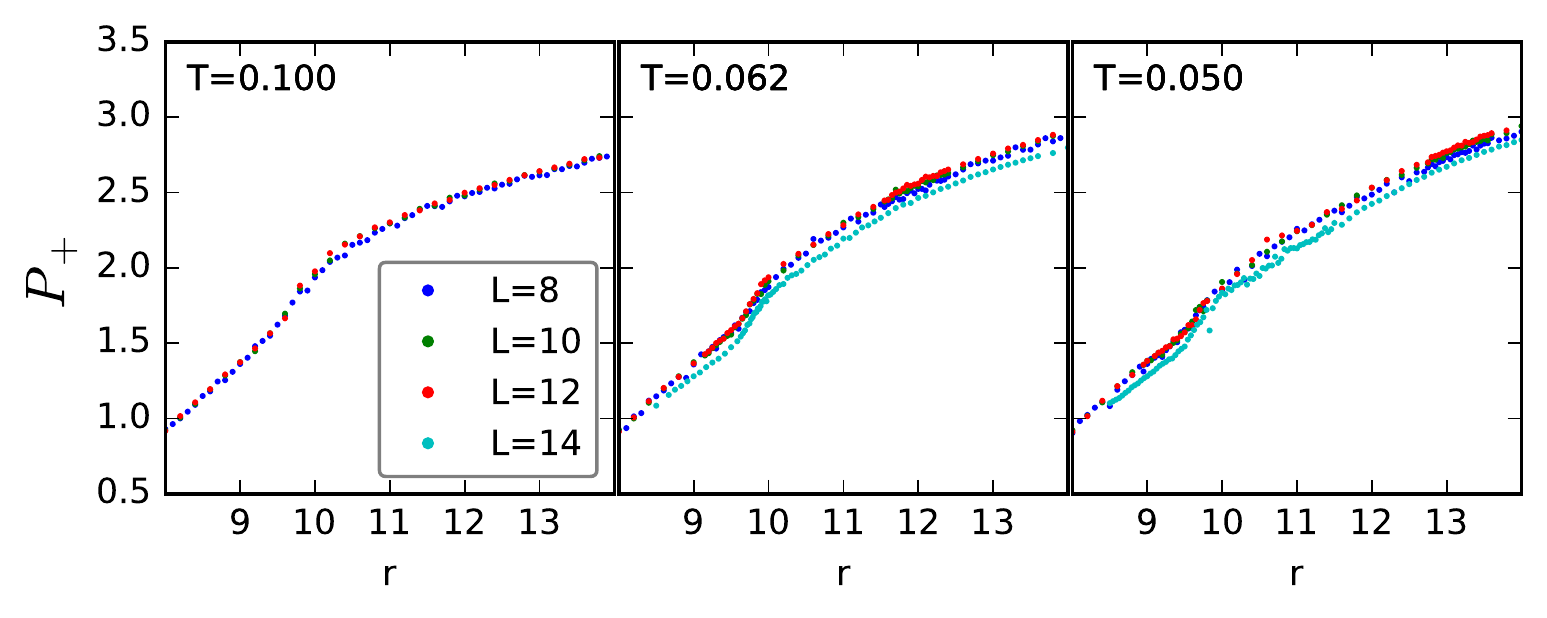}
  \caption{Superconducting susceptibility in the $s$--wave channel, $P_+$, across the phase diagram. Compare to the $d$--wave susceptibility shown in Fig.~\ref{fig:Pminus} in the main text.}
  \label{fig:P_of_r}
\end{figure}

At low temperatures $P_-$ is strongly dependent on $L$ (see Fig.~\ref{fig:Pminus} in the main text). In contrast, $P_+$ (shown in Fig.~\ref{fig:P_of_r}) remains size-independent. Note also that the $s$-wave susceptibility is smaller by more than two orders of magnitude than the $d$--wave one. While we have not attempted to extract the transition temperature from the finite size scaling behaviour of $P_-$, it is clear that the pairing instability occurs in the $d$-wave channel.

\section{Charge and pair density wave}
\begin{figure}[th]
  \centering
  \includegraphics[width=\linewidth]{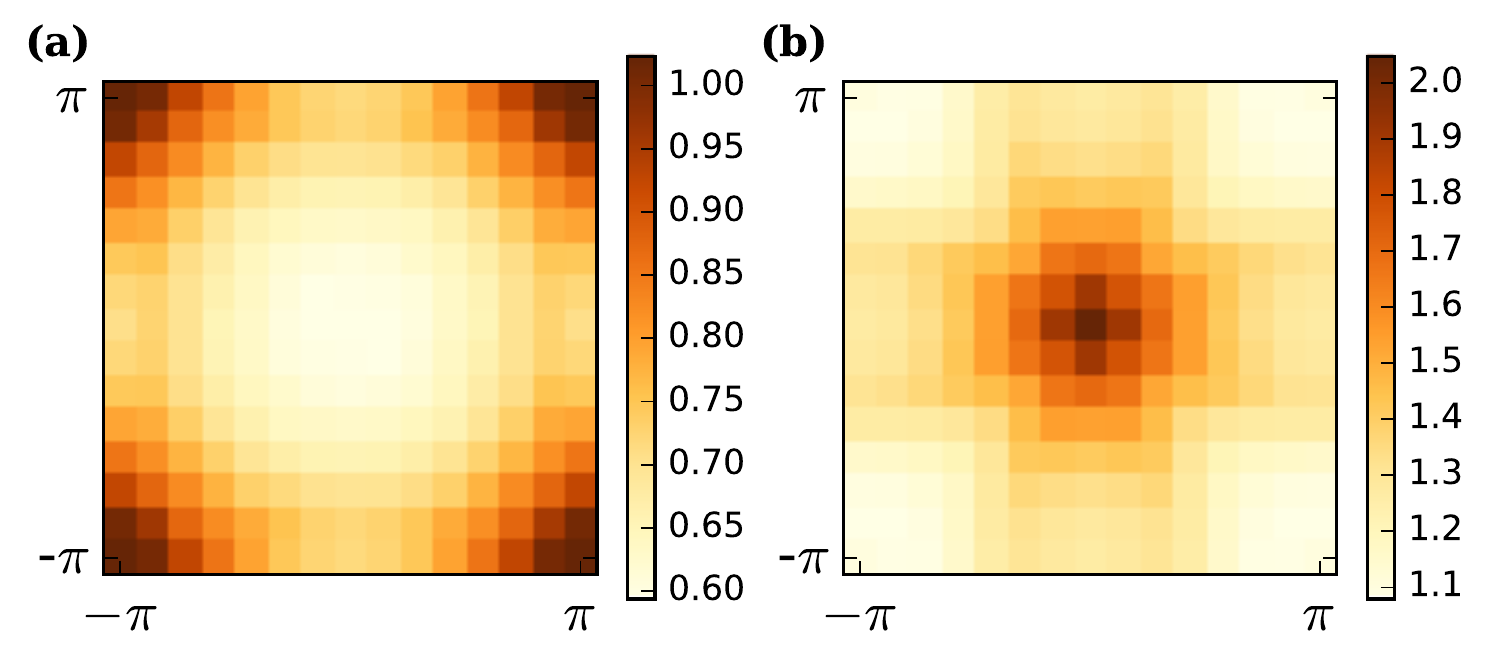}
  \caption{(a) $s$-wave CDW and (b) $s$-wave PDW susceptibilities, as defined in Eq.~\eqref{eq:susc}, across the Brillouin zone. 
  Shown here is data for $L=14$, $T=0.083$, and $r=10.4$.}
  \label{fig:s_wave_CDW_PDW_cmap}
\end{figure}

\begin{figure}[th]
  \centering
  \includegraphics[width=\linewidth]{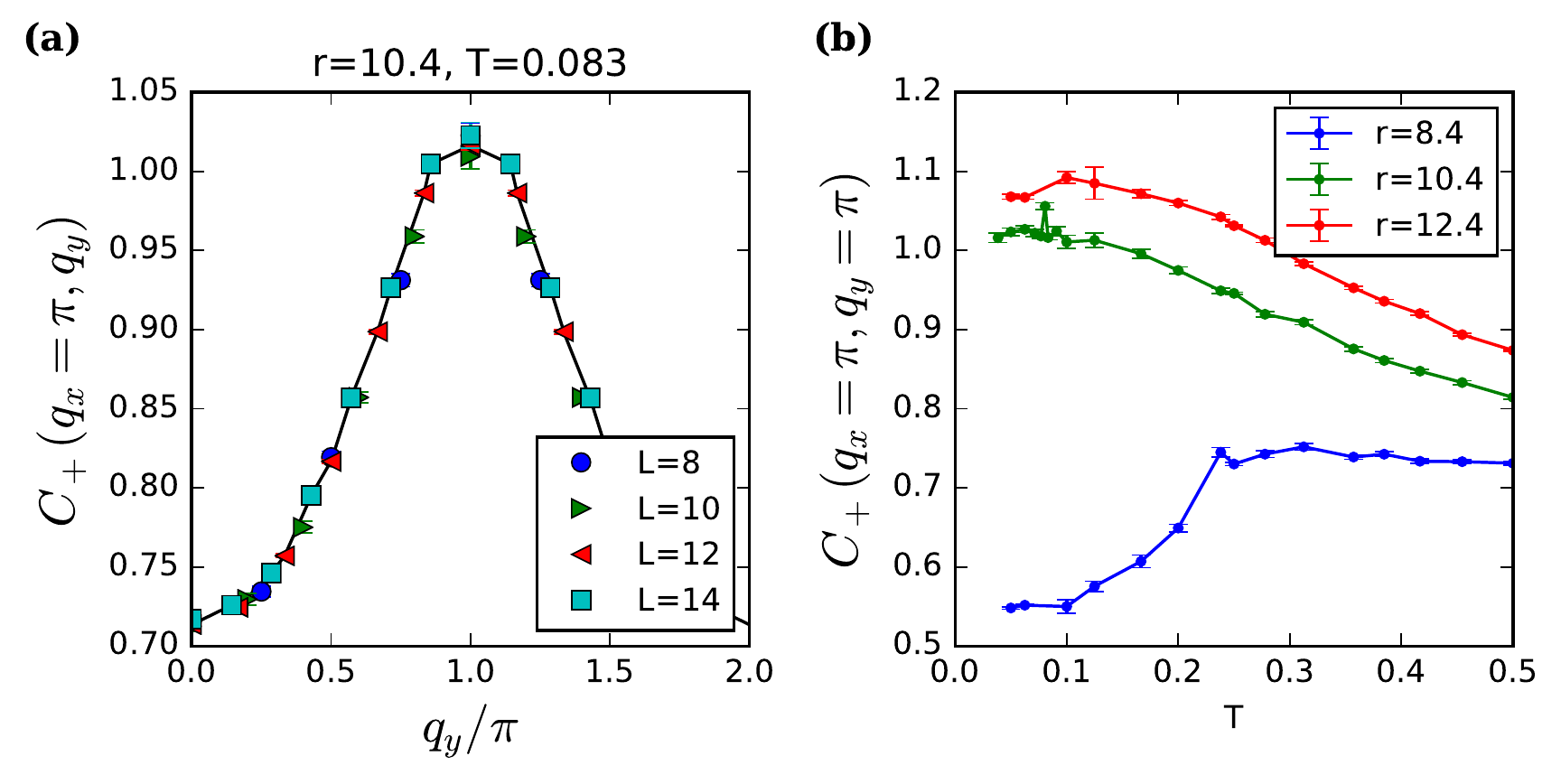}
  \caption{(a) The $s$-wave CDW susceptibility versus momentum along the high-symmetry cut $\mathbf q=(\pi,q_y)$ for various system sizes. 
  					The solid line is a guide to the eye. 
				     (b) Temperature dependence of the CDW susceptibility at $q=(\pi,\pi)$ for multiple values of $r$.}
  \label{fig:s_wave_CDW}
\end{figure}

In the main text we have focused on the $d$-wave CDW and PDW susceptibilities. The $s$-wave counterparts are shown in Fig.~\ref{fig:s_wave_CDW_PDW_cmap}. Much like $P_-$, $P_+$ shows no structure at finite momenta. $C_+$ is peaked close to $\mathbf q =(\pi,\pi)$ (see also Fig.~\ref{fig:s_wave_CDW}(a)), although the optimal $\mathbf{q}$ can vary slightly with $r$ (not shown). As the temperature is lowered, $C_+$ is at most moderately enhanced (see Fig.~\ref{fig:s_wave_CDW}(b)), and its maximal value decreases as with decreasing $r$.

\begin{figure}[th]
  \centering
  \includegraphics[width=\linewidth]{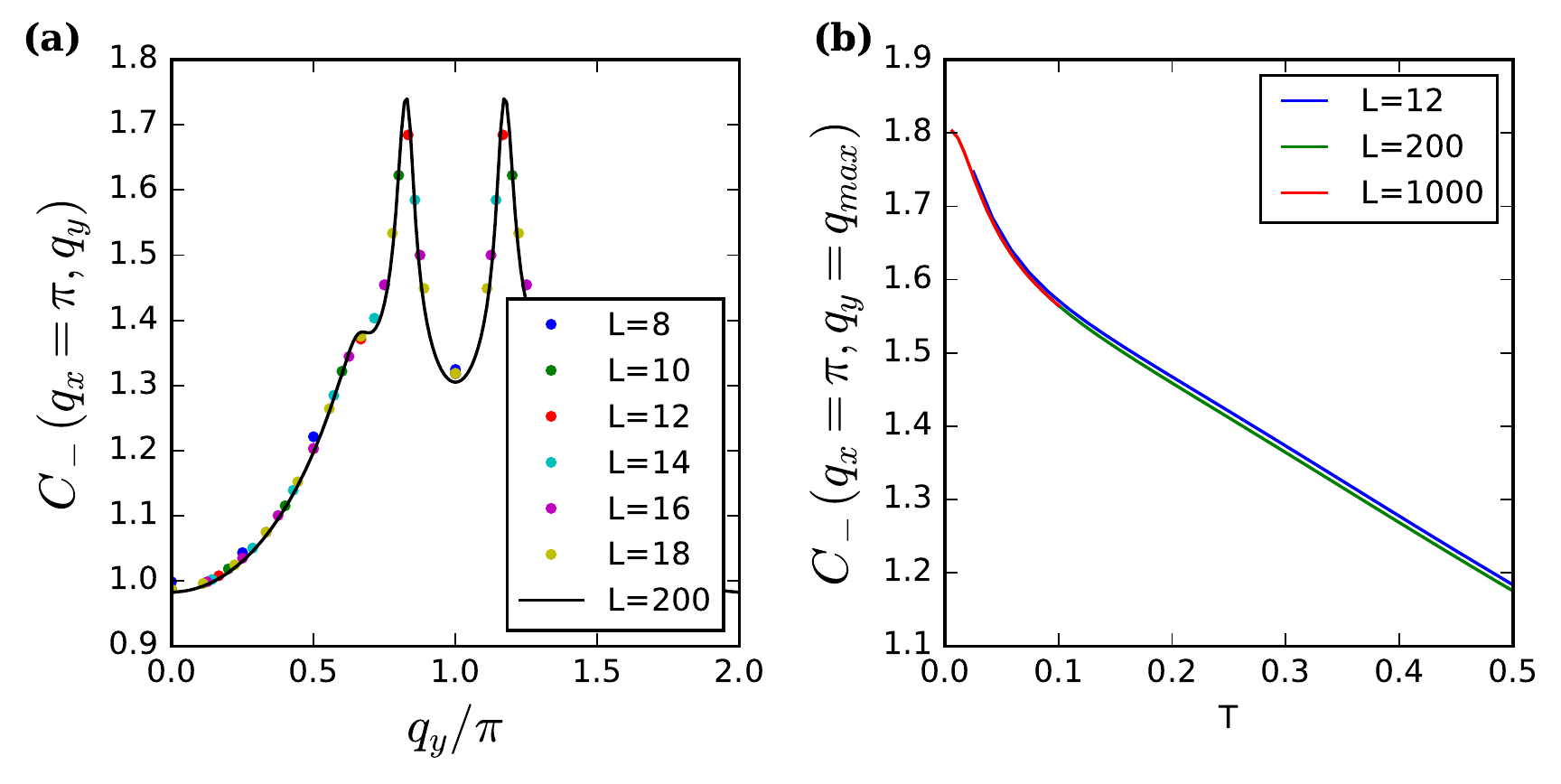}
  \caption{(a) The non-interacting CDW susceptibility versus momentum along the high-symmetry cut $\mathbf q=(\pi,q_y)$ for various system sizes, shown here at $T=0.025$ (b) Temperature dependence of the non-interacting CDW susceptibility}
  \label{fig:non-interacting_CDW}
\end{figure}

The quasi-one-dimensional character of the fermionic dispersion can account for the enhancement of the CDW susceptibility. In Figure~\ref{fig:non-interacting_CDW} we show the CDW susceptibility for the non-interacting ($\lambda=0$) case. Note that for this case $C_-(\mathbf q) = C_+(\mathbf q)$. At low temperatures, $C_-(\mathbf{q})$ is peaked at $\mathbf q=(\pi, q_{\max})=(\pi,0.92\pi)$, similar to the interacting model. As the temperature is lowered, $C_-(\pi,q_{\max})$  increases and saturates 
at low temperatures, see Fig.~\ref{fig:non-interacting_CDW}(b). Compared with $C_-$ in the interacting case, shown in Fig.~\ref{fig:cdw} of the main text, we see that the maximal CDW susceptibility in the interacting case is about 70\% larger than the non-interacting one. 

\end{document}